\documentclass[manuscript]{acmart}
\usepackage{framed}
\usepackage{url}
\usepackage{graphicx}
\usepackage{hyperref}
\usepackage{flushend}
\usepackage{cleveref}
\usepackage{amssymb}
\usepackage{amsmath}
\usepackage{xcolor}
\usepackage{soul}
\usepackage{algpseudocode}
\usepackage{algorithm}
\usepackage{xspace}
\usepackage{tabularx}
\usepackage{multirow}
\usepackage{balance}
\usepackage{subcaption}
\usepackage{booktabs} 
\usepackage{booktabs}

\PassOptionsToPackage{hyphens}{url}
\expandafter\def\expandafter\UrlBreaks\expandafter{\UrlBreaks
	\do\a\do\b\do\c\do\d\do\e\do\f\do\g\do\h\do\i\do\j%
	\do\k\do\l\do\m\do\n\do\o\do\p\do\q\do\r\do\s\do\t%
	\do\u\do\v\do\w\do\x\do\y\do\z\do\A\do\B\do\C\do\D%
	\do\E\do\F\do\G\do\H\do\I\do\J\do\K\do\L\do\M\do\N%
	\do\O\do\P\do\Q\do\R\do\S\do\T\do\U\do\V\do\W\do\X%
	\do\Y\do\Z}

\newcount\Comments  \Comments=1  
\newcommand{\kibitz}[2]{\ifnum\Comments=1\textcolor{#1}{#2}\fi}

\providecommand{\vs}{vs. }
\providecommand{\ie}{\emph{i.e.,} }
\providecommand{\eg}{\emph{e.g.,} }

\providecommand{\etal}{\emph{et al.\xspace}}
\providecommand{\etc}{\emph{etc.\xspace}}

\providecommand{\myparab}[1]{\smallskip\noindent\textbf{#1} }

\newcommand{\squishenum}{\begin{enumerate}{}{\setlength{\itemsep}{0pt}\setlength{\parsep}{0pt}\setlength{\topsep}{3pt}\setlength{\partopsep}{0pt}\setlength{\leftmargin}{1.5em}\setlength{\labelwidth}{1em}\setlength{\labelsep}{0.5em}}}
\newcommand{\squishlist}{\begin{list}{$\bullet$}{\setlength{\itemsep}{0pt}\setlength{\parsep}{3pt}\setlength{\topsep}{3pt}\setlength{\partopsep}{0pt}\setlength{\leftmargin}{1.5em}\setlength{\labelwidth}{1em}\setlength{\labelsep}{0.5em}}}
\newcommand{\squishlisttwo}{\begin{list}{$\bullet$}{\setlength{\itemsep}{0pt}\setlength{\parsep}{0pt}\setlength{\topsep}{0pt}\setlength{\partopsep}{0pt}\setlength{\leftmargin}{2em}\setlength{\labelwidth}{1.5em}\setlength{\labelsep}{0.5em}}}
\newcommand{\squishend}{\end{list}}
\newcommand{\squishenumend}{\end{enumerate}}

\setcopyright{none}
\begin{document}
\title{Online Political Discourse in the Trump Era}

\author{Rishab Nithyanand}
\affiliation{Data \& Society Research Institute}
\author{Brian Schaffner}
\affiliation{University of Massachusetts at Amherst}
\author{Phillipa Gill}
\affiliation{University of Massachusetts at Amherst}

\renewcommand{\shortauthors}{R. Nithyanand, et al.}

\begin{abstract}
We identify general trends in the (in)civility and complexity of political 
discussions occurring on Reddit between January 2007 and May 2017 -- a 
period spanning both terms of Barack Obama's presidency and the first 100 
days of Donald Trump's presidency. 

We then investigate four factors that are frequently hypothesized as having
contributed to the declining quality of American political discourse -- (1) the
rising popularity of Donald Trump, (2) increasing polarization and negative
partisanship, (3) the democratization of news media and the rise of fake news,
and (4) merging of fringe groups into mainstream political discussions.

\end{abstract}

%
%

\maketitle

\section{Introduction}

The 2016 election featured the two most disliked candidates in modern US
presidential election history competing in the context of decades of increasing
partisan polarization \cite{schaffner2017making}. In this paper we explore how
online political discourse during the election differed from discourse occurring
prior to it, in terms of incivility and linguistic complexity. We find that
incivility in online political discourse, even in non-partisan forums, is at an
all time high and linguistic complexity of discourse in partisan forums has
declined from a seventh-grade level to a first-grade level
(\Cref{sec:discourse}).

The election was noteworthy for the high levels of incivility and declining
complexity of discourse among political elites, particularly Donald Trump
\cite{schaffnertrump}. Research has shown that when people are exposed to
incivility from political elites that they themselves will respond by using more
offensive rhetoric \cite{gervais2014following, kwon2017aggression}. We explore
how Trump's increasing popularity impacted the civility and complexity of
discourse in partisan forums. Our work uncovers a strong correlation between
Trump's rise in popularity and the increasing incivility observed in Republican
forums on Reddit (\Cref{sec:trump-effect}). 

In may ways, the 2016 campaign was the logical culmination of two decades of
affective polarization that witnessed Democrats and Republicans grow
increasingly negative in their feelings about the opposing party. Political
scientists have documented the increasing polarization among Americans for quite
some time \cite{abramowitz2008polarization}; however, more recent work has
emphasized the emotion-based (affective) nature of this polarization. Drawing on
social identity theory \cite{tajfel1979integrative}, studies have found that one
of the defining features of partisan polarization is the increasingly negative
feelings that members of one party have for the other party
\cite{iyengar2012affect}. We measure the incidence of negative partisanship in
political forums and find a strong correlation with incivility, supporting the
theory that partisan identity leads people to experience emotions of both
enthusiasm and anger \cite{mason2016cross, huddy2015expressive}. Anger, in
particular, is likely to give rise to incivility due to its ability to motivate
political action \cite{groenendyk2014emotional, 
valentino2011election, huddy2015expressive}. Thus as Americans experience
political anger more frequently they are likely to be motivated to go online to
engage in political discussions \cite{ryan2012makes}. While we see that the 2016
election was not very dissimilar to 2012 (in terms of incidence of negative
partisanship), we find that negative partisanship has shown an upward trend even
after inauguration day (unlike 2012). We also find that hatred towards political
entities of both parties was at an all time high during the 2016 elections,
reinforcing the theory that 2016 was the ideal year for a non-establishment
candidate (\Cref{sec:polarization}). 

The 2016 campaign also witnessed unprecedented rhetoric from a major
presidential candidate regarding the credibility of the news media.
Additionally, during this time, public distrust of and anger at the
political establishment and traditional news media was at an all time high
\cite{gallup-media}. Taken together, these conditions can lead individuals to
engage in partisan motivated reasoning \cite{weeks2015emotions}, which can fuel
the spread and belief of ``fake news''. We explore how frequently misinformation
was shared and discussed online. We find that during the elections, Republican
forums shared and discussed articles from outlets known to spread conspiracy
theories, heavily biased news, and fake news at a rate 16 times higher than
prior to the election -- and more than any other time in the past decade. Our
study shows that this misinformation fuels the uncivil nature of discourse
(\Cref{sec:news}).

The racism (Trump's statements concerning Mexicans, Muslims, and other broad
groups), sexism (the Access Hollywood recordings), and general incivility
exhibited by the Trump campaign did not have any significant impact on his
presidential run. In fact, recent events (\eg Charlottesville and other Unite
the Right rallies) have shown that these actions have emboldened and brought
fringe groups into the mainstream. We investigate partisan forums and find a
significant overlap between participants in mainstream Republican and extremist
forums. We uncover a strong correlation between the rise in offensive discourse 
and discourse participation from extremists (\Cref{sec:fringe}).

\section{Reddit and the Reddit Dataset}
Reddit is the fourth most visited site in the United States and ninth most
visited site in the world \cite{Alexa-Reddit}.
At a high-level, Reddit is a social platform which enables
its users to post content to individual forums called
\emph{subreddits}. Reddit democratizes the creation and moderation of these
subreddits -- \ie any user may create a new subreddit and most content
moderation decisions are left to moderators chosen by the individual subreddit.
Subscribers of a subreddit are allowed to up-vote and down-vote posts made by
other users. These votes determine which posts are visible on the front
page of the subreddit (and, even the front-page of Reddit). 
Reddit also allows its users to discuss and have conversations about each post
through the use of comments. Specifically, subscribers of a subreddit can make
and also reply to comments on posts made within the subreddit. Like posts, the
comments may also be up-voted and down-voted. These votes determine which
comments are visible to users reading the discussion. 

Reddit is an attractive platform for analyzing political behaviour for three
main reasons: 
First, the democratization of content moderation and discussion combined with
the ability of participants to use pseudonymous identities has resulted in a
strong online disinhibition effect and
free-speech culture on Reddit \cite{Reddit-Freespeech}. This is unlike Facebook
which has stronger moderation policies and requires accounts to register with
their email addresses and real names (although the enforcement of both are
questionable).
Second, Reddit enables users to participate in long conversations and complex
discussions which are not limited by length. This is unlike Twitter which limits
posts and replies to 280 characters (prior to Sep 26, 2017 this limit was 140
characters \cite{TweetLength}).
Finally, Reddit allows scraping of its content and discussions. This has enabled
the community to build a dataset
\protect\footnote{\url{https://bigquery.cloud.google.com/dataset/fh-bigquery:reddit_comments}}
including every comment and post made since the site was made public in 2005. 

\begin{figure}[htb]
\centering
\includegraphics[trim=0cm 0cm 0cm 0cm, clip=true, width=.49\textwidth, angle=0]
{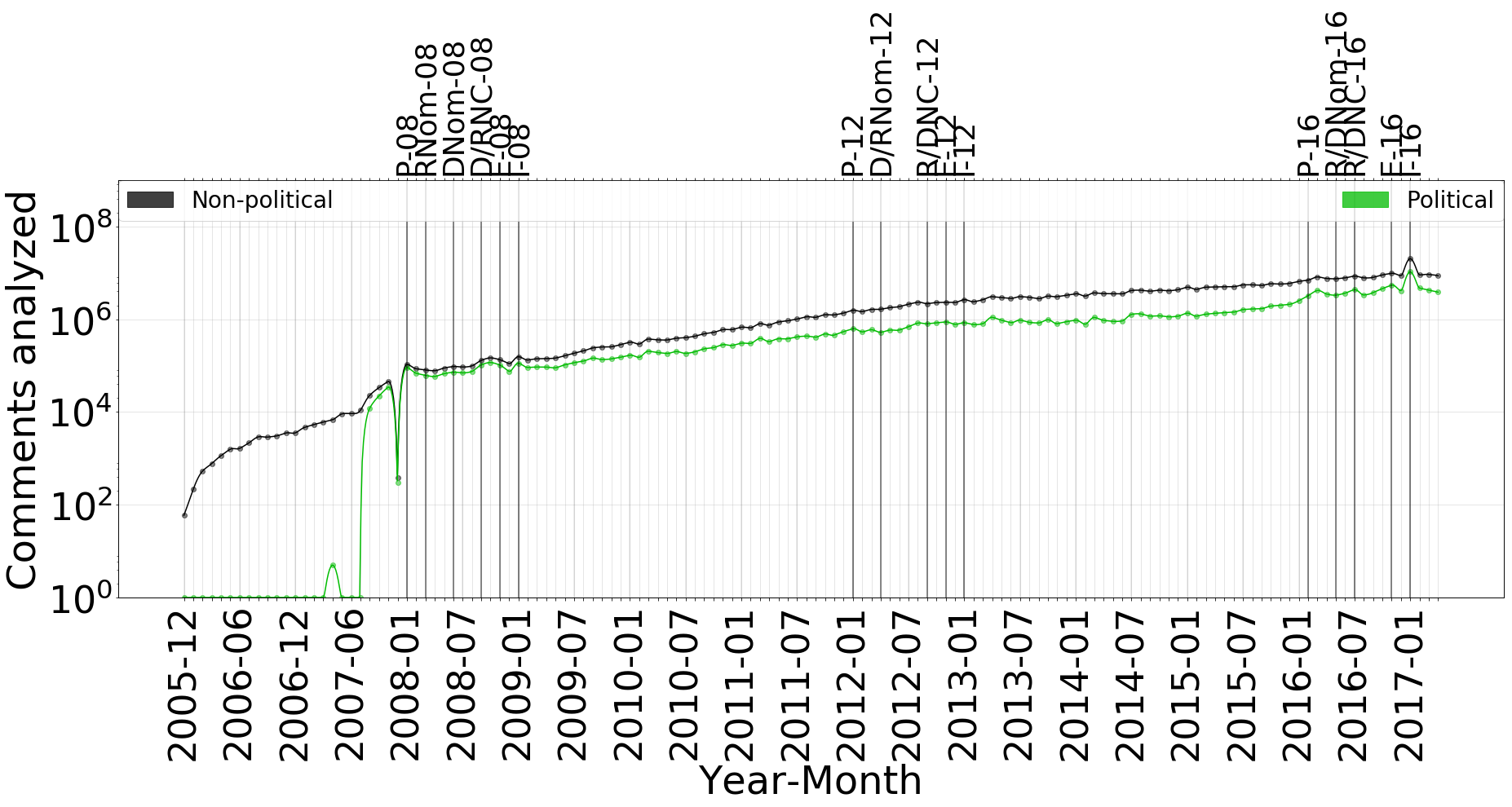}
\caption{\textbf{(log-scale)} Number of comments analyzed during each month from
December 2005 to June 2017. For each election year, P indicates the start of the
primaries, R/DNom indicates the month when the Republican/Democrat candidate
became the presumptive nominee, R/DNC indicates the month of the
Republican/Democratic National Conventions, E indicates the election month, and
I indicates the Presidential Inauguration.}
\label{fig:comments-analyzed}
\end{figure}

As of October 2017, the Reddit dataset includes a total of 3.5 billion comments
from 25.3 million authors made on 398 million posts. We categorize the posts and
comments in the dataset into two categories: political and non-political. Posts
and comments made in subreddits categorized by \emph{r/politics} moderators as
``related'' subreddits
\protect\footnote{https://www.reddit.com/r/politics/wiki/relatedsubs} are
tagged as political. We also tag the subreddits dedicated to all past
Democratic, Libertarian, and Republican presidential candidates as political.
All other subreddits are tagged as non-political. In total our political dataset
contained comments and posts from 124 subreddits -- each individually
categorized as general-interest, democratic, libertarian, republican,
international, and election-related.
In our study we focus on comments and posts made between December 1$^{st}$, 2005
and May 1$^{st}$ 2017 -- 100 days into Donald Trump's presidency. We analyze
every comment and post made in our set of political subreddits during this
period -- 130 million comments in 3 million posts -- and contrast these with a
random (10\%) sample of non-political comments made during the same period-- a
total of 332 million  comments in 12 million posts. \Cref{fig:comments-analyzed}
shows the number of political and non-political comments analyzed during each
month of from December 2005 to May 2017. It should be noted that the first
political subreddit appeared only in January 2007 -- therefore we have no
political content to analyze before this period.

\section{Civility and Complexity of Discourse}\label{sec:discourse}
In order to understand how online political discourse has evolved, we focus on
two concepts: (in)civility and complexity of discourse. 

\subsection{Incivility in political discourse}
We use the prevalence of offensive speech in political discussions on Reddit as
a metric for incivility. Previous work \cite{mutz2006hearing} has defined uncivil
discourse as ``communication that violates the norms of politeness'' -- a
definition that clearly includes offensive speech. 

\myparab{Identifying offensive speech.} 
In order to identify if a Reddit comment contains offensive speech, we make use
of the offensive speech classifier proposed by Nithyanand \etal
\cite{Nithyanand-FOCI2017}. At a high-level, the classifier uses a Random Forest
model built upon the cosine similarities between a ``hate vector'' and annotated
training data, both embedded within a 100-dimensional word embedding constructed
from every Reddit comment. The approach yields an accuracy between 89-96\% on
testing data. The complete specification and evaluation are described in
\cite{Nithyanand-FOCI2017}. We note that the classifier is unable to
differentiate between offensive comments and comments which quote offensive
content -- \eg comments quoting Donald Trump's candidacy announcement speech,
which included derogatory remarks about Mexican immigrants
\cite{Trump-announcement}, were also classified as offensive. To identify the
entities in offensive comments, we use the SpaCy \cite{spacy} entity recognition
toolkit augmented with a custom dictionary of political entities.

\begin{figure*}[htb]
\centering
\begin{subfigure}[b]{.485\textwidth}
\includegraphics[trim=0cm 0cm 0cm 0cm, clip=true, width=\textwidth, angle=0]
{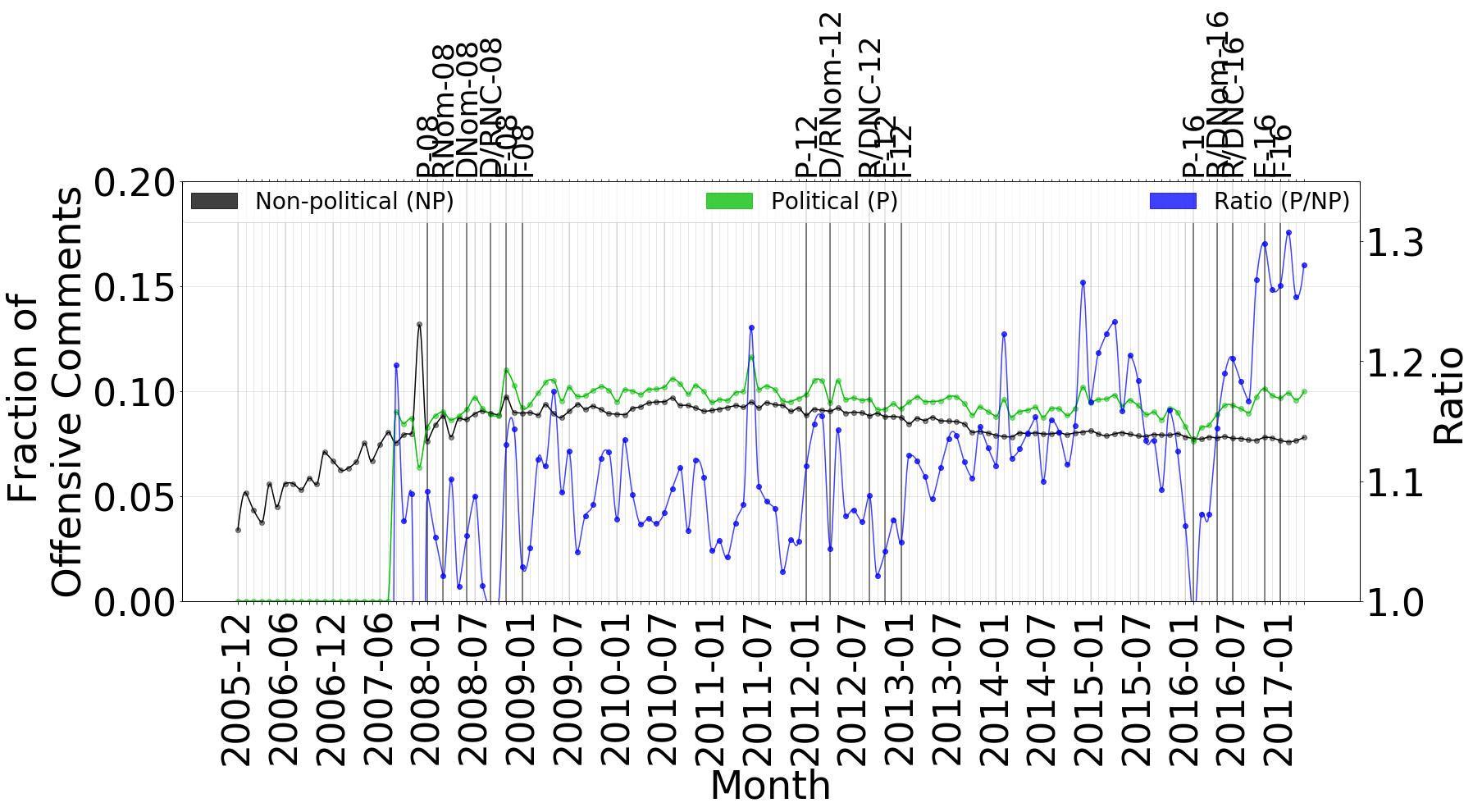}
\caption{Political \vs Non-political subreddits: Fraction of offensive comments.}
\label{fig:offensiveness-comment-trends-pol-apol}
\end{subfigure}
\begin{subfigure}[b]{.485\textwidth}
\includegraphics[trim=0cm 0cm 0cm 0cm, clip=true, width=\textwidth, angle=0]
{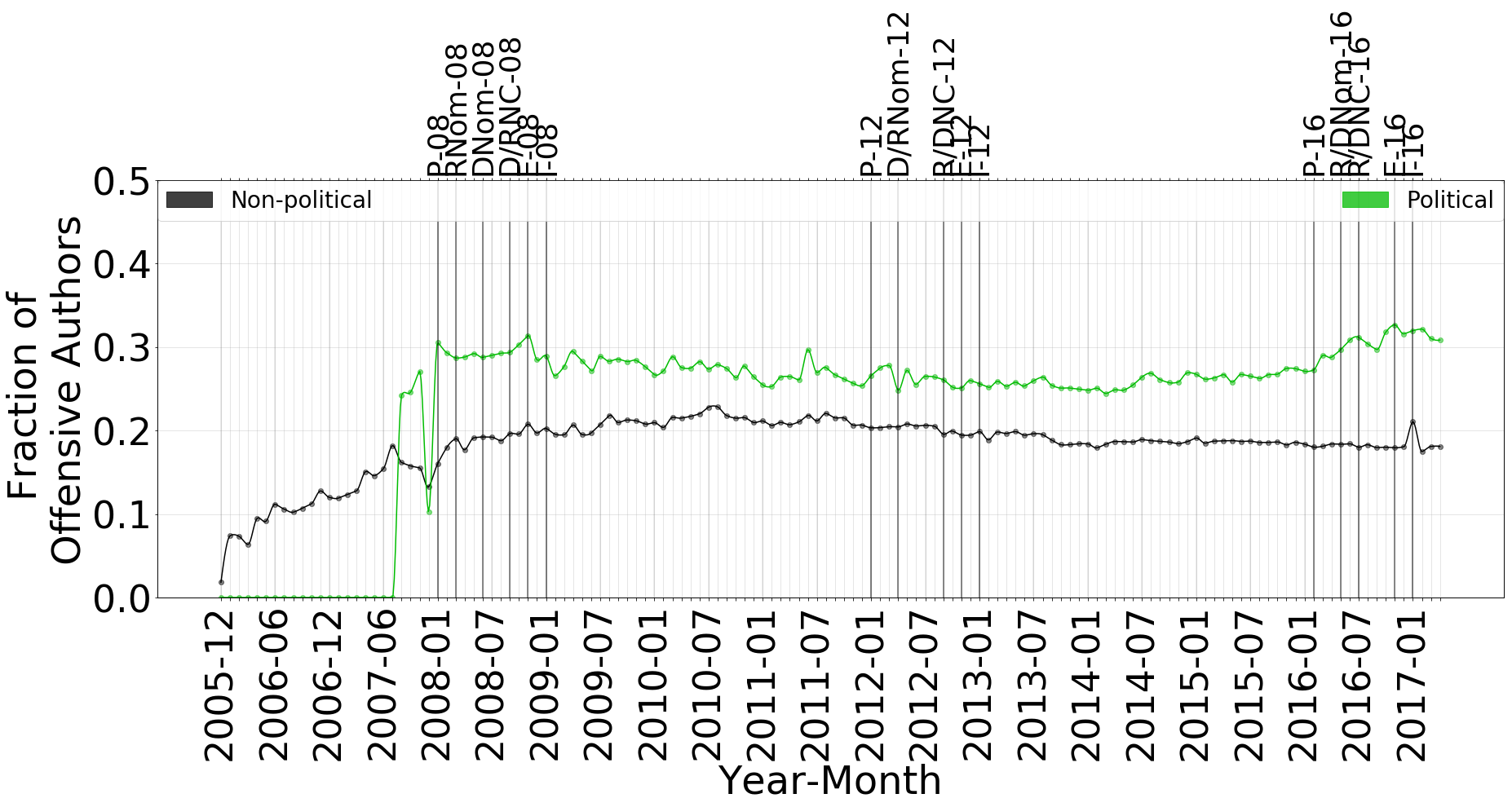}
\caption{Political \vs Non-political subreddits: Fraction of authors posting at least one offensive comment.}
\label{fig:offensiveness-author-trends-pol-apol}
\end{subfigure}

\begin{subfigure}[b]{.485\textwidth}
\includegraphics[trim=0cm 0cm 0cm 0cm, clip=true, width=\textwidth, angle=0]
{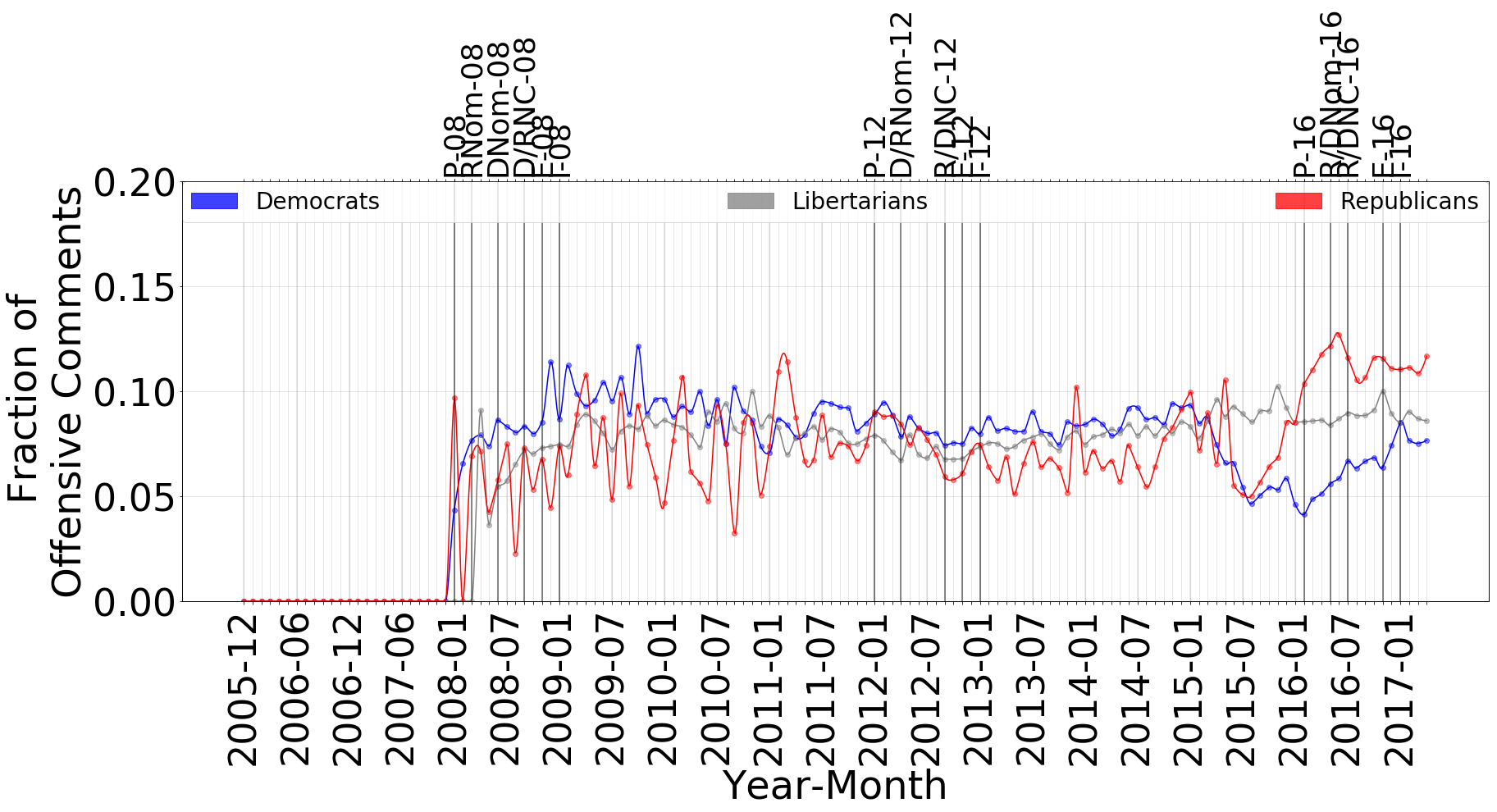}
\caption{Partisan subreddits: Fraction of offensive comments.}
\label{fig:offensiveness-comment-trends-parties}
\end{subfigure}
\begin{subfigure}[b]{.485\textwidth}
\includegraphics[trim=0cm 0cm 0cm 0cm, clip=true, width=\textwidth, angle=0]
{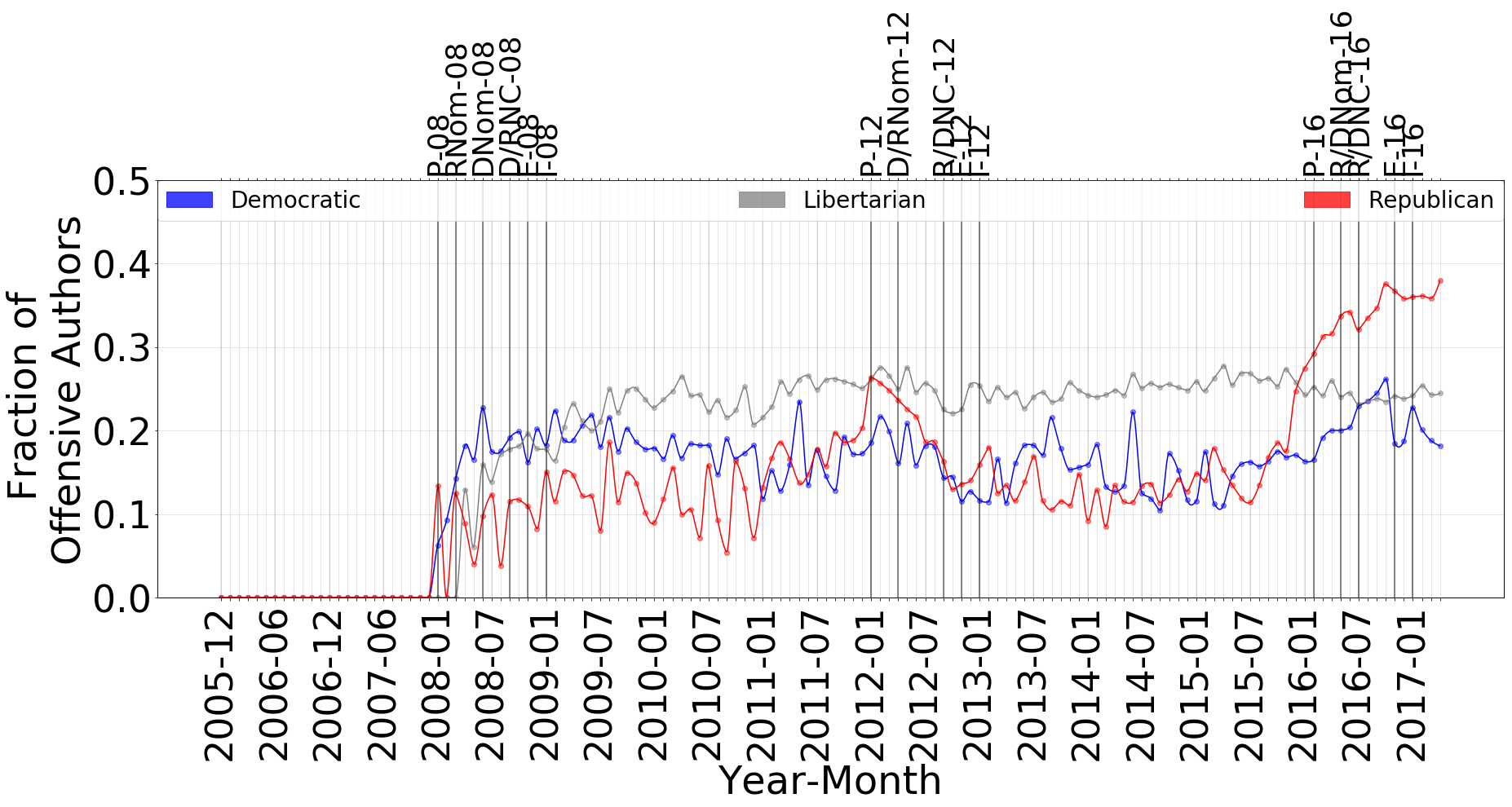}
\caption{Partisan subreddits: Fraction of authors posting at least one offensive comment.}
\label{fig:offensiveness-author-trends-parties}
\end{subfigure}
\caption{Incidence of offensiveness in comments and the fraction of offensive
authors for each month.}
\label{fig:offensiveness-trends-pol-apol}
\end{figure*}

\myparab{Trends in offensive political discourse.}
\Cref{fig:offensiveness-comment-trends-pol-apol} shows how the incidence of
offensiveness has changed over time for subreddits in our political and
non-political datasets. We find that offensive comments in political subreddits
have always been at least as frequently occurring as offensive comments in
non-political subreddits. \Cref{fig:offensiveness-author-trends-pol-apol} shows
the fraction of all authors that posted at least one offensive comment during
each month. We find that authors of comments in political subreddits are much
more (nearly 35\%, on average) likely to be offensive than authors not
participating in political discussions.

Our data shows that the difference in incidence rates of offensive comments
between political and non-political subreddits has dramatically increased since
the start of the 2016 US presidential elections. In fact, we see that prior to
2014, there is only one month -- June 2011, during the debt-ceiling crisis in
congress and after Obama's announcement to withdraw large numbers of American
forces from Afghanistan -- where political comments were over 20\% more likely
to be offensive than non-political comments. 
Since then, we notice this to be true for short periods of time in 2014 and
2015, and for the entire period from July 2016 until May 2017. Inspecting the
offensive comments made during these periods, we find that large fractions (over
35\%) of offensive comments were targeted at law enforcement authorities and the
Black Lives Matter movement for the events surrounding the deaths of James Boyd
(2014), Michael Brown (2014), and Freddie Gray (2015).
The increase in incivility of discourse since July 2016 is attributed to the
start of the US Presidential elections and the conclusions of the Democratic and
Republican National Conventions -- with over 80\% of all offensive comments
targeted at the two political parties and politicians including Hillary Clinton,
Bernie Sanders, and Donald Trump. Worryingly, even after the elections and
inauguration, the incidence of offensiveness in political comments and the
fraction of offensive political comment authors has continued to grow. As of
May 2017, we find that (1) approximately 10\% of all political comments are
classified as offensive, nearly 30\% higher than for non-political comments and
(2) nearly one-third of all political comment authors made offensive comments,
over 70\% higher than for non-political comment authors.
 
\emph{Take-away:} Our results show that political discourse from May
2016 to May 2017 has been more offensive (and by our definition, uncivil) than
any other 12-month period in Reddit's 12 year history.
 
\myparab{Subreddits responsible for offensive political discourse.}
\Cref{fig:offensiveness-comment-trends-parties} shows how the incidence of
offensiveness has changed over time in subreddits categorized as Democratic,
Libertarian, and Republican. We find several interesting long-term trends --
until 2015 the comments on Democratic subreddits were on average 23\% and 15\%
more likely to be offensive than comments on Republican and Libertarian
subreddits, respectively. However, since 2015, comments on Republican subreddits
were on average 46\% and 7\% more offensive than Democratic and Libertarian
subreddits. We find similar trends in
\Cref{fig:offensiveness-author-trends-parties} which shows the fraction of all
authors that posted at least one offensive comment in a Democratic, Libertarian,
and Republic subreddit during each month. The incidence of offensiveness in
Libertarian subreddits on the other hand remains fairly stable through the
entire period of the study with only one spike over the 10\% mark in June 2015
-- the month Donald Trump announced his candidacy.

Looking closer at specific events responsible for spikes in offensive discourse
reveals that prior to the start of the 2016 election season, comments in the
Republican subreddits were most offensive (12\% incidence rate) during early
2011 and 2014 -- the period during Barack Obama's 2011/2014 State of the Union
addresses and the attempts to repeal (2011) and expand (2014) the Affordable Care
Act. We see a large spike in the incidence of offensive comments starting from
Donald Trump's candidacy announcement in June 2015 (5.1\% of comments and 12\%
of authors) to Trump's victory of the Republican nomination in May 2016 (12.8\%
of comments and 35\% of authors). Further, in spite of a drop in incidence of
offensiveness in comments to 11.6\% after the elections, the fraction of
offensive comment authors has continued to grow to 38\% as of May 2017. 
On the Democratic side, 2015 was the least offensive period in Democratic
subreddits with incidence of offensive comments varying between 7\% and 4.8\%.
Further, despite the growing rate of offensiveness during the 2016 primaries and
general election -- peaking between the election in November 2016 (6.3\% of
comments) and inauguration in January 2017 (8.5\% of comments), this period
remained the least offensive election cycle in Democratic subreddits -- even
compared to 2012 when Barack Obama was uncontested in the primaries. It is
interesting to note that in spite of the low incidence of offensiveness, this
period saw the highest number of offensive comment authors in the Democratic
subreddits -- peaking at 25\% in October 2016.

\emph{Takeaway:} Offensive political discourse has grown at a high rate in
Republican subreddits. As of May 2017, comments in Republican
subreddits were 55\% more likely to be offensive than comments in Democratic
subreddits and with nearly twice as many authors of offensive comments.



\begin{figure*}[htb]
\centering
\begin{subfigure}[b]{.485\textwidth}
\includegraphics[trim=0cm 0cm 0cm 0cm, clip=true, width=\textwidth, angle=0]
{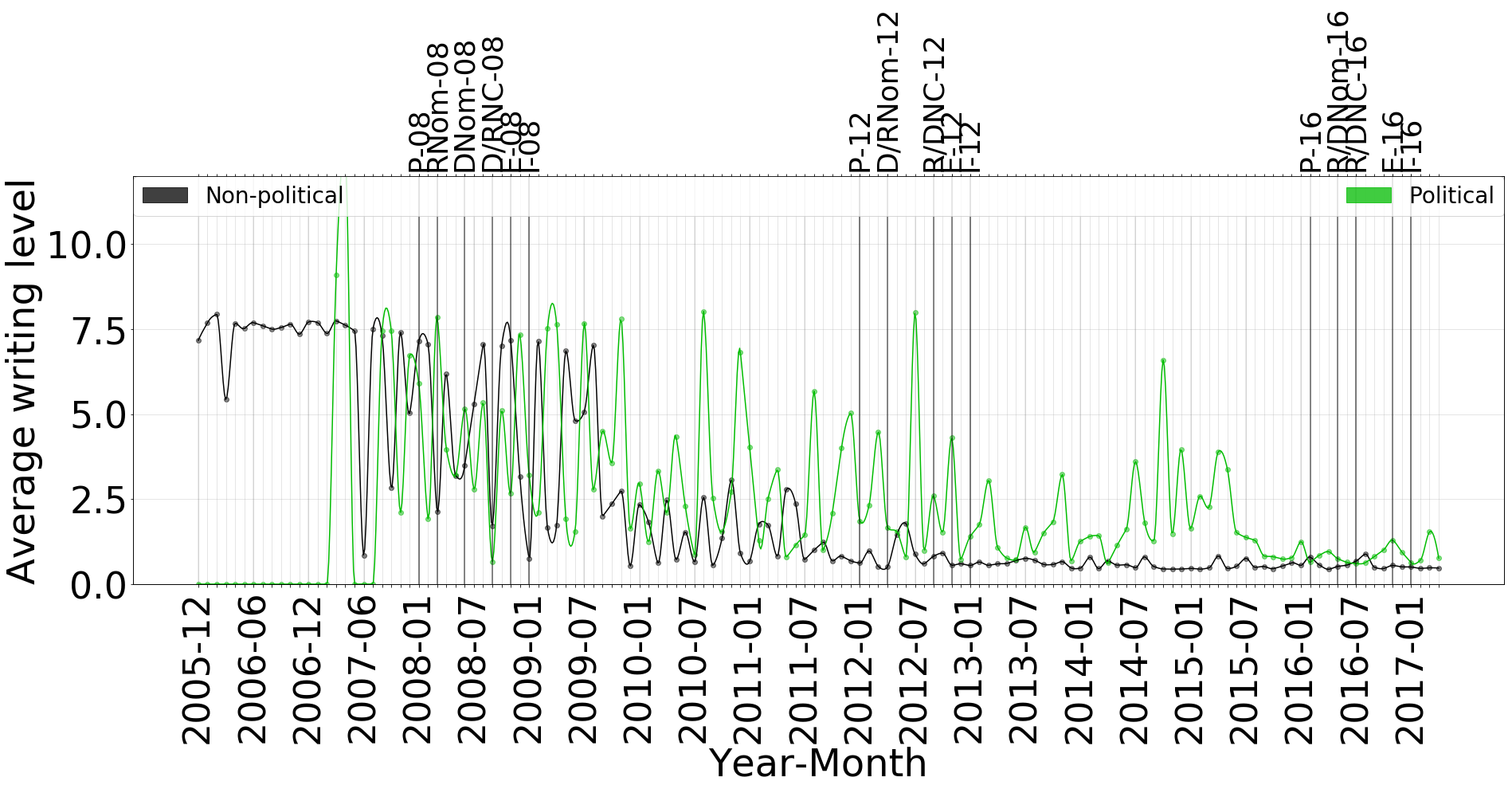}
\caption{Political \vs Non-political subreddits.}
\label{fig:writing-levels-fk}
\end{subfigure}
\begin{subfigure}[b]{.485\textwidth}
\includegraphics[trim=0cm 0cm 0cm 0cm, clip=true, width=\textwidth, angle=0]
{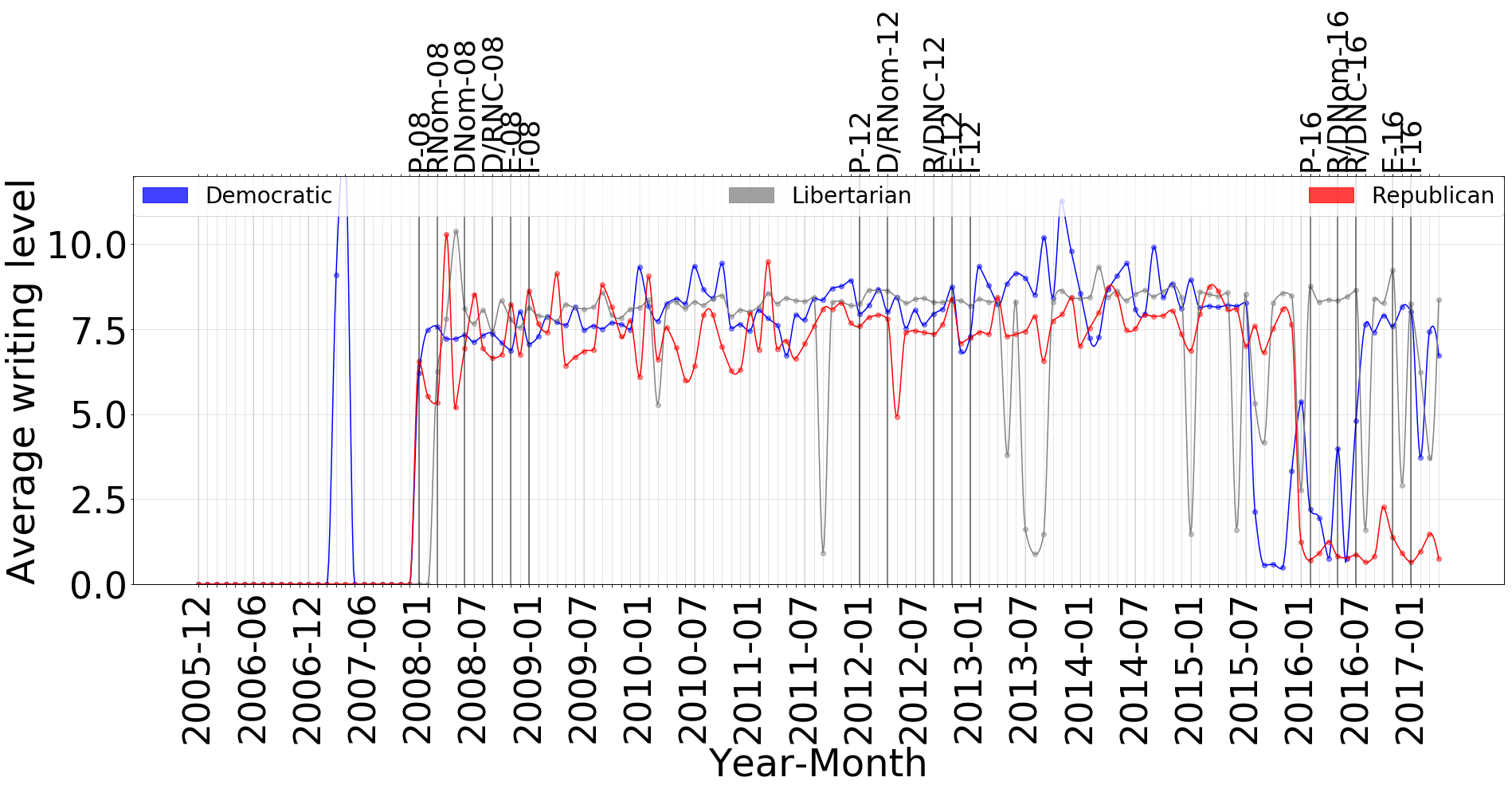}
\caption{Partisan subreddits.}
\label{fig:writing-levels-fk-party}
\end{subfigure}
\caption{Average Flesch-Kincaid grade level of comments.}
\label{fig:flesch-kincaid}
\end{figure*}

\subsection{Complexity of political discourse}
We focus on linguistic complexity and use the Flesch-Kincaid readability
grade-level \cite{Flesch-Kincaid} as a metric. The Flesch-Kincaid metric assigns
higher scores to text containing longer words and sentences
(\Cref{eq:flesch-kincaid}) -- which generally tend to be more complex. This
approach has been used in the past to understand the complexity of political
speeches and is used in government and military documents in the United States.
\begin{equation}
\label{eq:flesch-kincaid}
Grade = 0.39 \times \frac{words}{sentences} + 11.8 \times \frac{syllables}{words} - 15.59
\end{equation}
\myparab{Trends in linguistic complexity of discourse.}
\Cref{fig:flesch-kincaid} shows the linguistic complexity of comments made for
each month in political and non-political subreddits
(\Cref{fig:writing-levels-fk}) and also broken down by Democratic, Libertarian,
and Republican subreddits (\Cref{fig:writing-levels-fk-party}). We see that
discourse in political subreddits is generally more complex than in
non-political subreddits, despite being highly variable over time. Deeper
analysis shows that this variability is introduced by inclusion of the large
``general-interest'' political subreddit communities (\eg \emph{r/politics} and
\emph{r/worldnews}) which have over 1 million comment authors.  

Considering only the partisan subreddits (\Cref{fig:writing-levels-fk-party}),
we see that comments had an average readability
grade-level between 7.8 (Democratic subreddits) and 7.5 (Republican and
Libertarian subreddits) until December 2015, with only marginal variations
throughout. During the 2016 Democratic and Republican primaries (January - June
2016), however, there were significant drops in complexity -- Democratic and
Republican subreddits had an average reading grade-level of 2.6 and 1.9,
respectively. Complexity of discourse on Libertarian subreddits, on the other
hand, improved to a 7.6 grade. These results suggest that the highly contested
intra-party primaries on both sides led to much lower quality of discourse even
on partisan subreddits. Since the end of the primaries (June 2016 - May
2017), complexity of discourse in Democratic subreddits improved to a 6.9
grade-level while discourse in Republican subreddits further declined to a 1.1
grade-level. During this same time, discourse in Libertarian subreddits also
slightly declined to a 6.6 grade-level. 

\emph{Takeaway:} The complexity of discourse in partisan subreddits was at its
historical lowest during the 2016 primaries and presidential elections. While
the complexity of discourse has recovered in the Democratic subreddits since the
election, it has continued to decline to a first grade-level in
Republican subreddits. 
 

\section{The Trump Effect}\label{sec:trump-effect}
Anecdotal evidence has suggested that the rise in Donald Trump's popularity
resulted in more offensive political discourse. This has been referred to as the
``Trump Effect'' \cite{trump-effect}. Since we cannot prove or disprove the
causal nature of the Trump Effect, we instead study the linear correlation
between Donald Trump's popularity and the offensiveness and complexity of
political discourse (measured in \Cref{sec:discourse}).

As a metric for Trump's popularity, we use poll data aggregated by Real Clear
Politics during the 2016 elections \cite{rcp-polls} and approval/disapproval
data aggregated by 538 since the start of the Trump presidency \cite{538-polls}.
We split the poll data from Real Clear Politics into two categories:
primary and general-election related polls. From the period between Trump's
candidacy announcement speech and his clinching of the Republican nomination
(June 2015 - May 2016), we only focus on his weekly average vote share in polls
related to the Republican primaries. Similarly, from July 2016 (the conclusion of
the Democratic and Republican National Conventions) until November 8,
2016 (Election day) we only focus on Trump's weekly average vote share in polls
related to the general election and for the period following the presidential
inauguration (Jan 2017 - May 2017), we only focus on Trump's average approval
and disapproval ratings as reported by 538.

\myparab{The Republican primaries (June 2015 - May 2016).}
During the primaries, we find that Trump's rise in popularity was strongly
positively correlated with the rise of offensive discourse in Republican
subreddits (Pearson correlation co-efficient: .84, p-value < .0001) and strongly
negatively correlated with the complexity of discourse in Republican subreddits 
(Pearson correlation co-efficient: -.65, p-value < .0001). We do not find
statistically significant correlations between Trump's rise in popularity and
political discourse in the Democratic or Libertarian subreddits.

\myparab{The general election (July 2016 - November 2016).}
Trump's popularity during the general election did not have a
significant correlation with the complexity of discourse in any subreddits.
However, his popularity was moderately correlated with offensiveness in
Democratic subreddits (Pearson correlation co-efficient: .49, p-value < .005).
Interestingly, Hillary Clinton's popularity during this period was
also moderately correlated with the offensiveness in Republican subreddits
(Pearson correlation co-efficient: .54, p-value < .005). This points to the
change in the nature of discourse from intra- to inter-party elections --
\ie that offensive discourse in inter-party elections are correlated to the
success of the ``other''. This supports recent scholarship noting the rise
of negative partisanship and the fact that individuals are generally motivated
to engage in political discourse due to anger with the opposition
\cite{huddy2015expressive}.

\myparab{Donald Trump's Presidency (January 2017 - May 2017).}
During the first 100 days of Trump's presidency, we find that there is only a
statistically significant  correlation between his approval (and disapproval)
ratings and the offensiveness in Republican subreddits. As was the case during
the general elections, there is no statistically significant correlation
between Trump's popularity and complexity of discourse. We find a moderate
negative correlation between Trump's approval rating and offensive discourse in
Republican subreddits (Pearson correlation co-efficient: -.59, p-value: < .05)
and a moderate positive correlation between Trump's disapproval rating and
offensiveness in Republican subreddits (Pearson correlation co-efficient: .55,
p-value < .05). It is unclear if this rise in offensiveness occurs due to
attempts to ``double down'' in support of Trump or due to displeasure with the
course of Trump's presidency.  

\emph{Takeaway:} We find that Donald Trump's popularity is always, at least
moderately, correlated with the offensiveness of political discourse. During the
primaries, Trump's popularity was strongly correlated with the rise of
offensiveness in Republican subreddits. During the general election, Trump's
popularity was moderately correlated with offensiveness in Democratic
subreddits and during his presidency, there is a moderate negative  correlation
between his approval ratings and offensiveness in Republican subreddits. 
 
\section{Negative Partisanship}\label{sec:polarization}

Recent work \cite{negative-partisanship} has suggested that
``persistent and durable repulsion from a political party'', defined as
\emph{negative partisanship}, has an effect on voting decisions and election
turnout. We explore the incidence of negative partisanship on
Reddit and seek to understand how it relates to the decline of civility and
complexity of discourse. 

We use two metrics as a measure of negative partisanship: (1) the fraction of
political comments in a partisan subreddit that express strong negative
sentiments towards the opposition party -- \eg fraction of all comments in the
Democratic subreddits which express negative sentiments towards the Republican
party, and (2) the number of political entities that are most commonly featured
in comments classified as offensive (\ie considering all subreddits). While the
first metric captures the traditional definition of negative partisanship, the
second captures the trend of a growing hatred towards all political entities
(or, the establishment). 

We use NLTK's Vader
\cite{sentiment-vader} sentiment analysis method to identify the sentiment of
a comment. Vader returns a compound sentiment score in the [-1, +1] range, where
-1 is the most negative sentiment and +1 is the most positive sentiment. We only
consider comments with a compound sentiment $\le-.70$ -- \ie strongly negative
comments. To identify political
entities in comments, we use the SpaCy entity recognition method \cite{spacy}
with a custom dictionary of political entities (manually curated from the common
nouns that occur close to the words "Democrats", "Republicans", and
"Libertarians" in our Reddit word embedding). When there are multiple political
entities in a comment, it is unclear how to properly associate the sentiment of
the comment with each entity -- \ie our sentiment analysis is at the
comment-level, not entity-level -- therefore we discard these comments. The same
approach is used to identify entities that are the targets of offensive
comments.

\begin{figure}[htb]
\centering
\includegraphics[trim=0cm 0cm 0cm 0cm, clip=true, width=.49\textwidth, angle=0]
{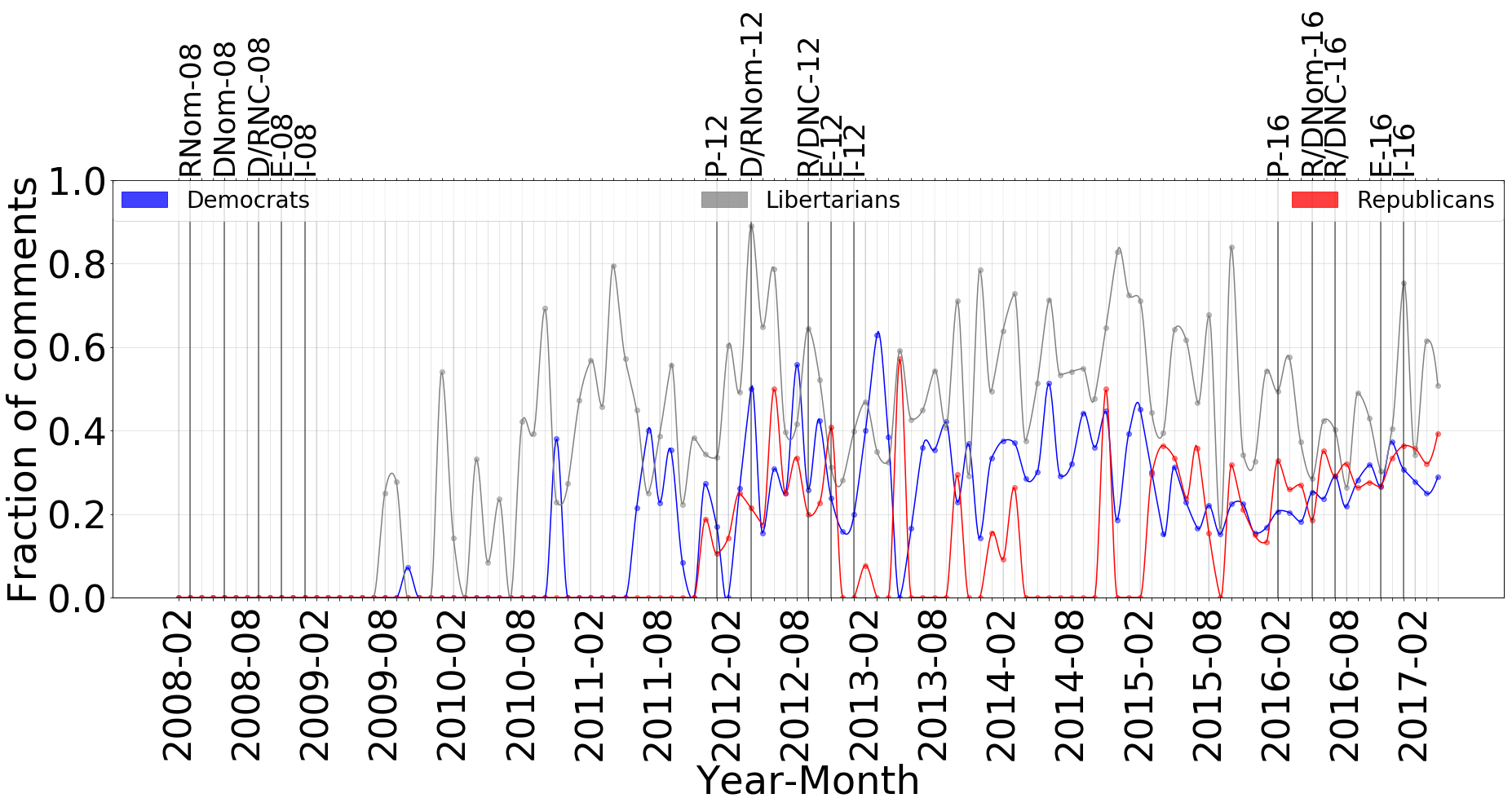}
\caption{Fraction of comments referencing the opposition party that have strong
negative sentiments.}
\label{fig:negative-partisanship}
\end{figure}

\Cref{fig:negative-partisanship} shows the fraction of comments referencing 
opposition parties that have strong negative sentiments (\vs comments that
refer to opposition parties and have other sentiments). We find that
Libertarians are most likely to refer to the Democratic and Republican party 
with strong negative sentiments -- on average over 45\% of all references to
these parties is strongly negative and only 7\% are positive. While the
Democratic subreddits have generally expressed negative sentiments against
opposition parties -- the trend declined in the period prior to and during the
early phase of Democratic primaries, suggesting that intra-party elections shift
the focus away from the inter-party dynamics. Between Super Tuesday III (April
2016) and Election night, negative partisanship on the Democratic subreddits
nearly doubled from 19\% to 37\%. We see a similar trend in the Republican
subreddits. This is possibly explained by the fact that Hillary Clinton and
Donald Trump had all but clinched their parties nominations after Super Tuesday
III and focus of their supporters was shifted to the general election. Between
Trump's clinching of the nomination and May 2017, negative partisanship on the
Republican subreddits grew from 20\% to 39\% (a 30-month high) displaying signs
of continuing the upward trend. In contrast, since the conclusion of the 2016
elections, negative partisanship on the Democratic subreddits declined to
28\% in May 2017. When considering only data since June 2015 -- the start of the
primary campaign season, we find that there are statistically significant
correlations between the incidence of negative partisanship and the decline of
civility in political discourse, suggesting that incivility in political
subreddits is frequently targeted at opposition parties. The observed
correlation is found to be much stronger in Democratic subreddits (Pearson
correlation co-efficient: .75, p-value < .0005) than in Republican subreddits
(Pearson correlation co-efficient: .39, p-value < .001). We also find a
moderately negative correlation between complexity of discourse on Republican
subreddits and the incidence of negative partisanship (Pearson correlation
co-efficient: -.40, p-value < .01).

\begin{figure}[htb]
\centering
\includegraphics[trim=0cm 0cm 0cm 0cm, clip=true, width=.49\textwidth, angle=0]
{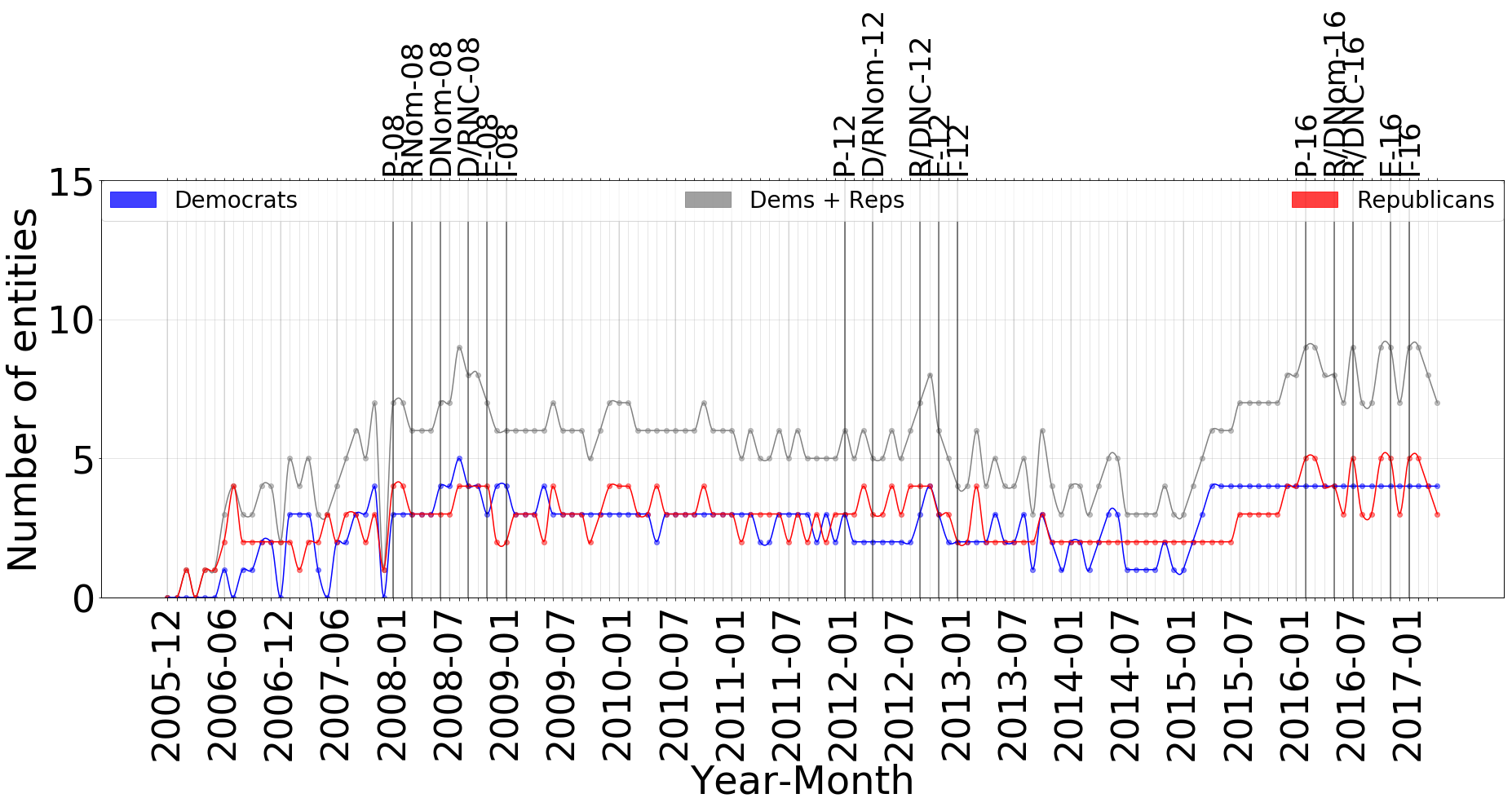}
\caption{Number of Democratic and Republican entities in Reddit's 100 most
commonly offended entities.}
\label{fig:targets-of-offensiveness}
\end{figure}

To gain a general sense of how political entities are viewed by Reddit (all
subreddits, including non-political), we ranked (all) entities by the number of
times they were the sole entity in a comment classified as offensive. The
results are illustrated in \Cref{fig:targets-of-offensiveness}. We find that
political entities have always been amongst Reddit's top 100 most offended
entities since 2006, peaking during the 2008, 2012, and 2016 presidential
elections. The number of political entities making an appearance in the 100 most
offended entities was at an all time high (of nine entities) during the months
leading up to the 2016 elections. Interestingly, we also find that the sitting
President and other ``establishment'' figures such as the speaker and majority
leader always rank in the Top 20 most offended entities.

\emph{Takeaway:} Although negative partisanship was at a 30-month high on
Republican subreddits, it was comparable to the 2012 election season. However,
the hatred shown towards specific political ``establishment'' entities was
unprecedented -- suggesting that 2016 was indeed the year of the outsider.

\section{Fake News and the Democratization of Media}\label{sec:news}
In this section we explore the impact of news media consumption habits on the
quality of political discourse. Specifically we focus on the impact of media
from controversial outlets (known for peddling conspiracy theories, \etc)
and democratized social platforms (YouTube and Twitter) that are increasingly
being repurposed for dissemination of ``news''.

\myparab{Rise of controversial media outlets.}
In our study we focus on the impact of conspiracy theory peddling, heavily
biased, fake, and foreign state-sponsored news outlets on political discourse on
Reddit. We use tags assigned by the OpenSources project
\protect\footnote{http://www.opensources.co/} to
identify when a news outlet falls in the above categories. We broadly categorize
these outlets as \emph{controversial}. We observe that of the 833 outlets
identified by the OpenSources project, 487 domains were active prior
to May 2015, 219 domains made their first appearance on Reddit after June 2015,
and 127 domains did not appear on Reddit. 

\begin{figure*}[htb]
\centering
\begin{subfigure}[b]{.49\textwidth}
\includegraphics[trim=0cm 0cm 0cm 0cm, clip=true, width=\textwidth, angle=0]
{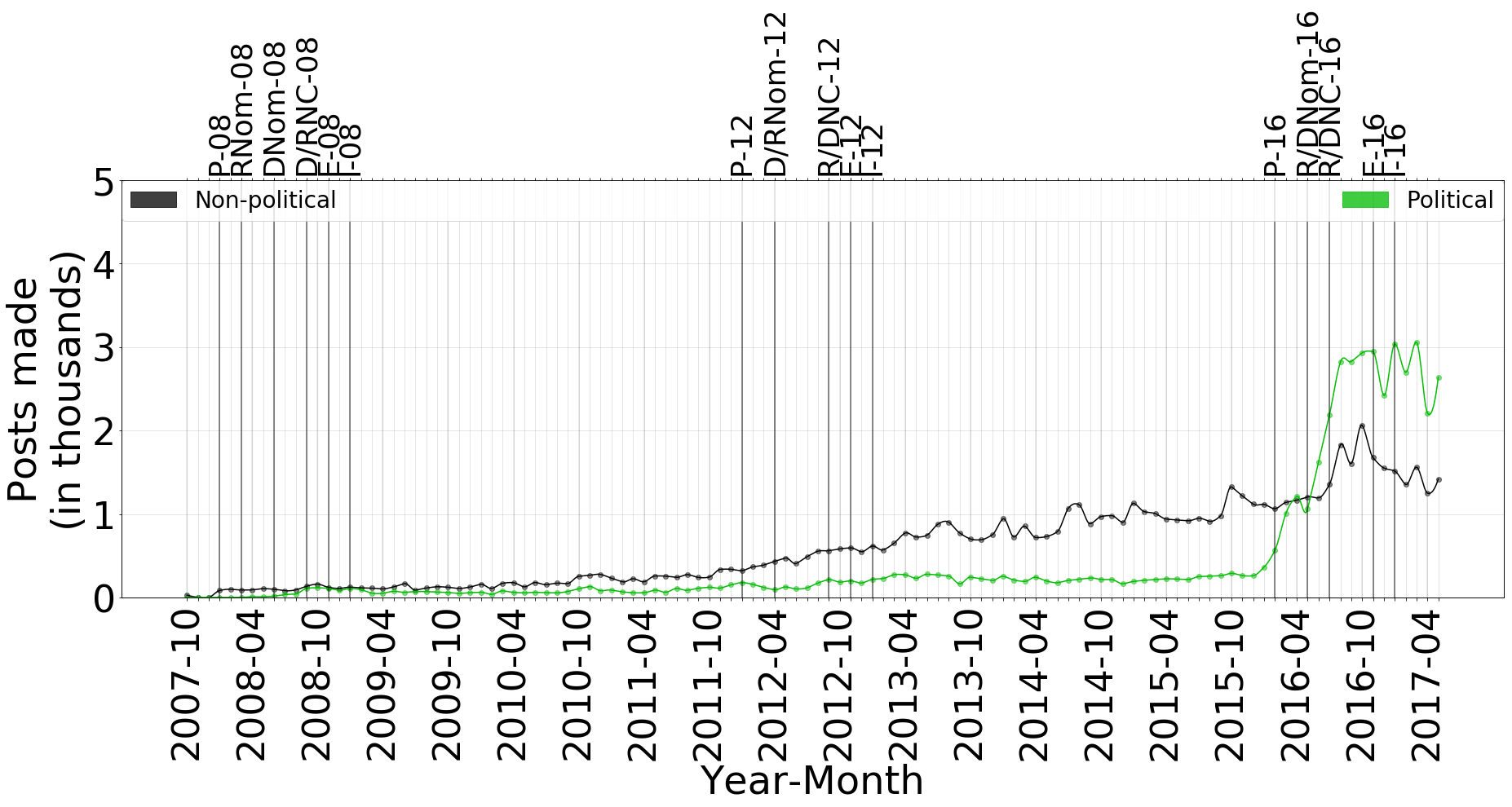}
\caption{Political \vs Non-political subreddits: Number of posts linking to
controversial outlets.}
\label{fig:conspiracy:posts:political-apolitical}
\end{subfigure}
\begin{subfigure}[b]{.49\textwidth}
\includegraphics[trim=0cm 0cm 0cm 0cm, clip=true, width=\textwidth, angle=0]
{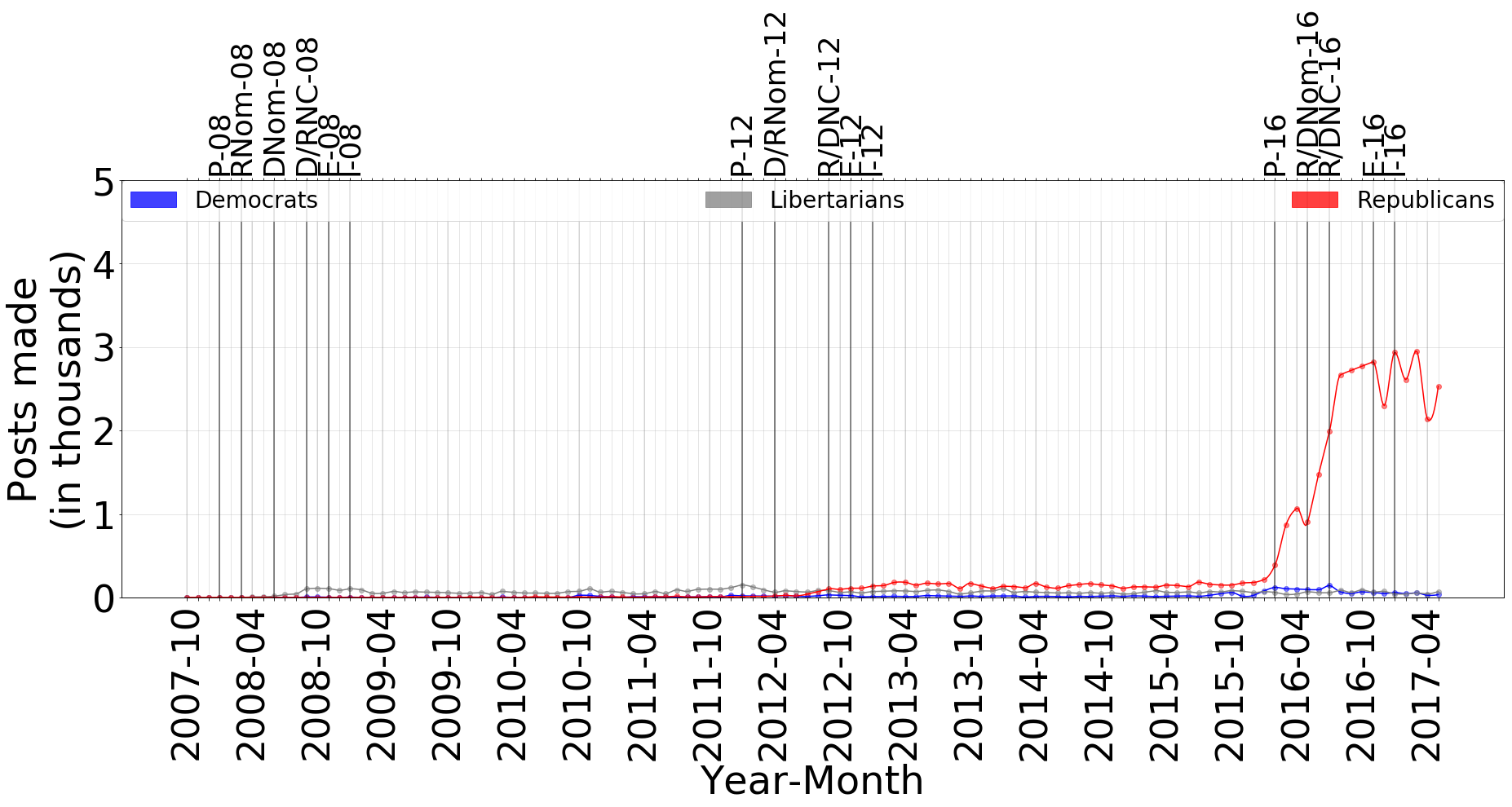}
\caption{Partisan subreddits: Number of posts linking to controversial outlets.}
\label{fig:conspiracy:posts:partisan}
\end{subfigure}

\begin{subfigure}[b]{.49\textwidth}
\includegraphics[trim=0cm 0cm 0cm 0cm, clip=true, width=\textwidth, angle=0]
{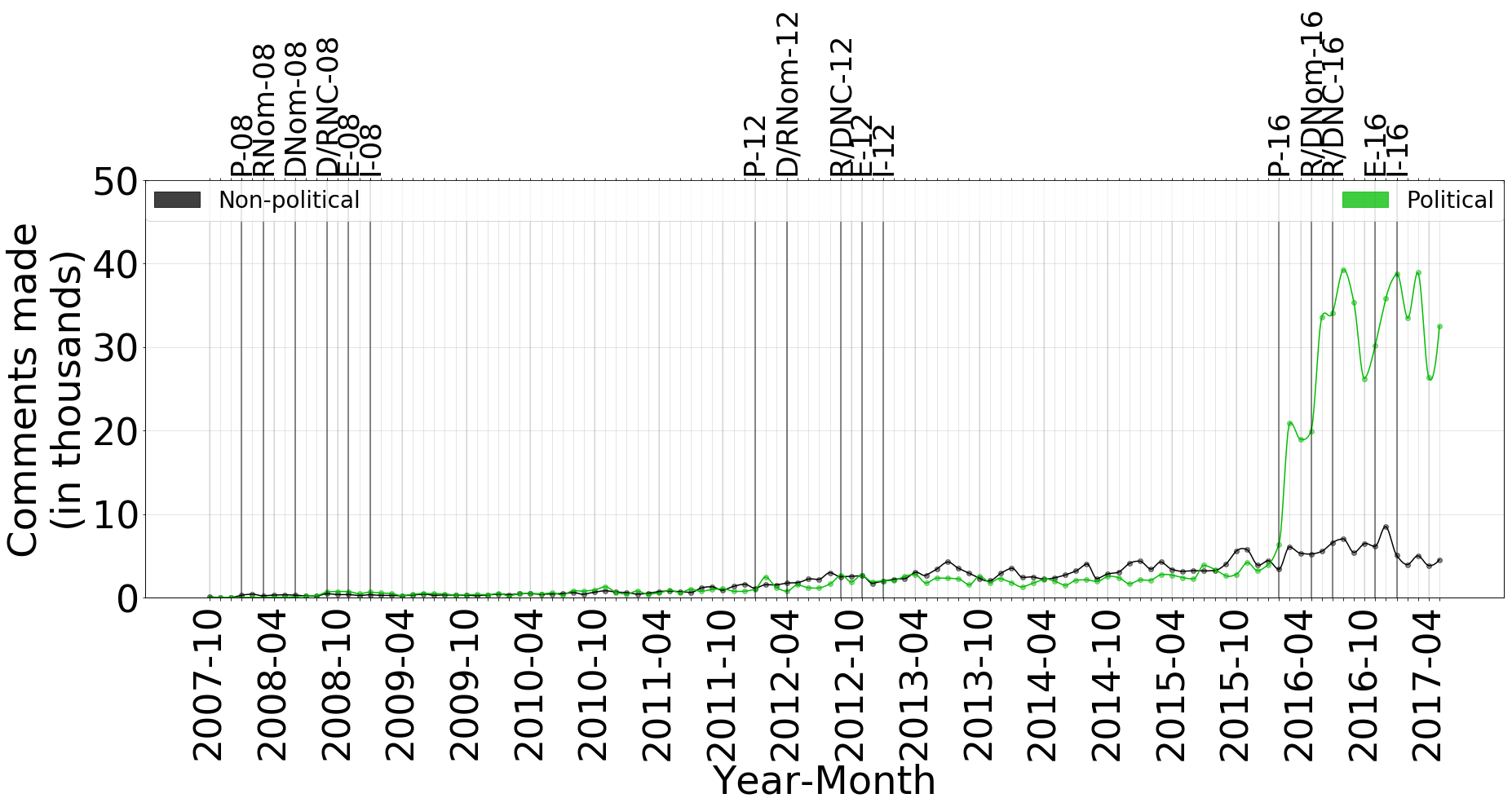}
\caption{Political \vs Non-political subreddits: Number of comments made in
posts from controversial outlets.}
\label{fig:conspiracy:comments:political-apolitical}
\end{subfigure}
\begin{subfigure}[b]{.49\textwidth}
\includegraphics[trim=0cm 0cm 0cm 0cm, clip=true, width=\textwidth, angle=0]
{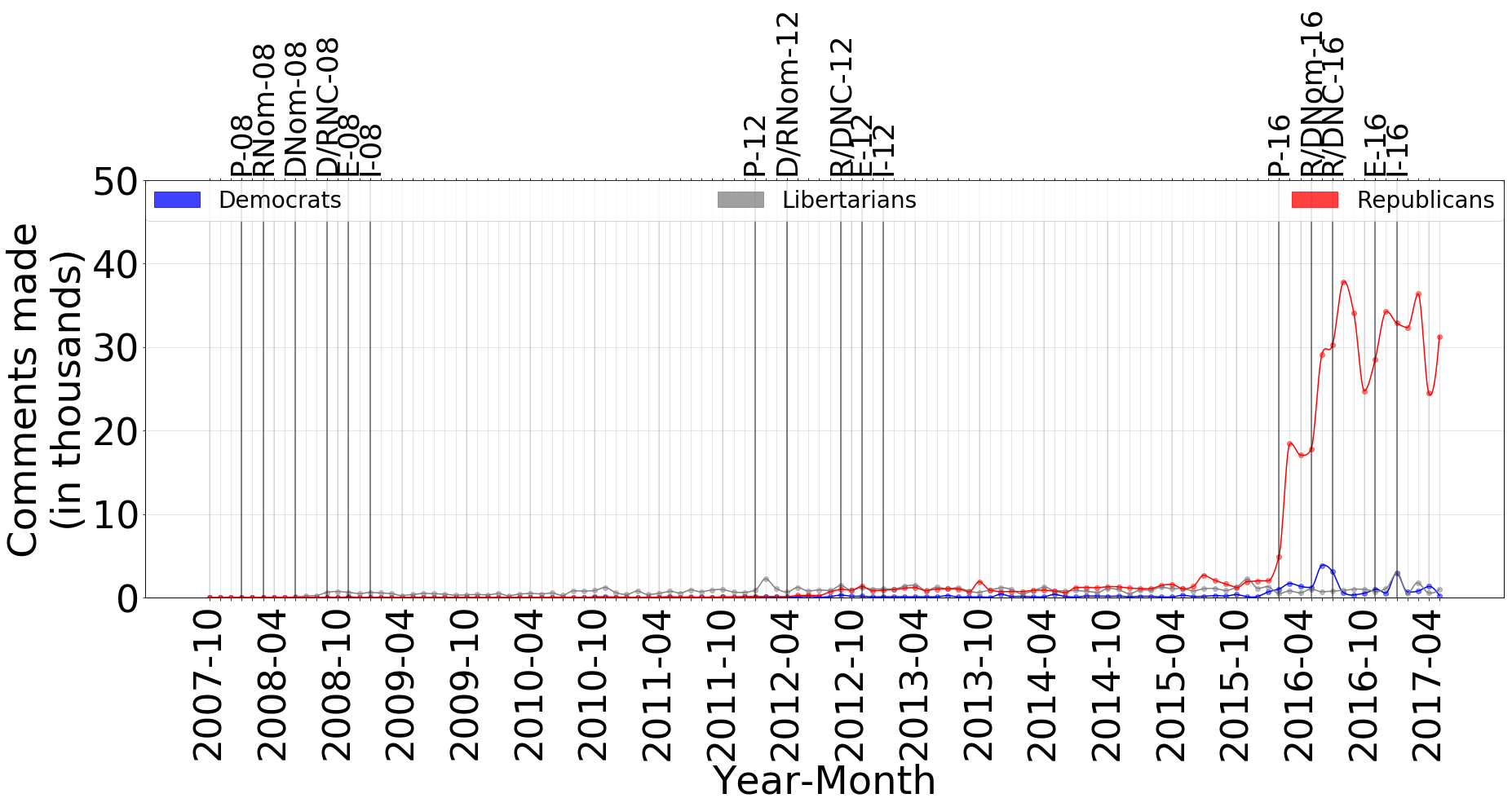}
\caption{Partisan subreddits: Number of comments made in posts from
controversial outlets.}
\label{fig:conspiracy:posts:partisan}
\end{subfigure}

\caption{Activity surrounding controversial media outlets on Reddit.}
\label{fig:controversial-domains}
\end{figure*}

\Cref{fig:controversial-domains} shows the amount of activity (in terms of
posts and comments) surrounding all controversial outlets. We find
that Republican subreddits were orders of magnitude more likely to be exposed
to articles associated with these outlets than any other group -- accounting
for over 80\% of all posting and commenting activity on links to controversial
outlets, during and after the 2016 election cycle. Interestingly, we see that
this was not the case prior to the elections. Links to controversial media
outlets were up to 600\% and 1600\% more likely during the Republican primaries
and the general election than in the months prior to the start of the 2015
Republican primaries. Since the start of Trump's presidency, the activity
surrounding links to controversial outlets continues to remain high. Upon
further investigation, we find that the subreddits \emph{r/The\_Donald} and
\emph{r/conservative} were the most commonly targeted subreddits. Although we do
not perform a thorough investigation of this anomalous behaviour in this paper,
we use this as evidence in our ongoing investigation of a coordinated
misinformation campaign targeted at Republican subreddits. 

In general (across all political subreddits), the incidence of offensiveness is
nearly 30\% higher in comments associated with controversial posts (compared to
all non-controversial posts). This provides a possible explanation for why
discourse was much more offensive in Republican subreddits. This hypothesis is
supported by a reasonably strong positive and statistically significant
correlation between the incidence of controversial posts and fraction of
offensive comments (Pearson correlation co-efficient: .59, p-value < .0001). We
do not find statistically significant correlations between the complexity of
discourse in Republican subreddits and incidence of posts from controversial
outlets, however. In the Democratic subreddits, we find that a majority of posts
(64\%) from controversial outlets had no comment activity -- suggesting that
these were removed by subreddit moderators or ignored by the community. There
were no statistically significant correlations between the incidence of
controversial posts and political discourse in the Democratic subreddits.

\emph{Takeaway:} Republican subreddits experienced a 1600\% increase in links to
controversial media outlets during the general elections. Combined with
the inflammatory nature typical of these articles, this offers an explanation
for the drastic growth of offensiveness in Republican subreddits. In Democratic
subreddits, there is little to no activity on posts from controversial media
outlets, suggesting more effective moderation and community policing.

\begin{figure}[htb]
\centering
\begin{subfigure}[b]{.49\textwidth}
\includegraphics[trim=0cm 0cm 0cm 0cm, clip=true, width=\textwidth, angle=0]
{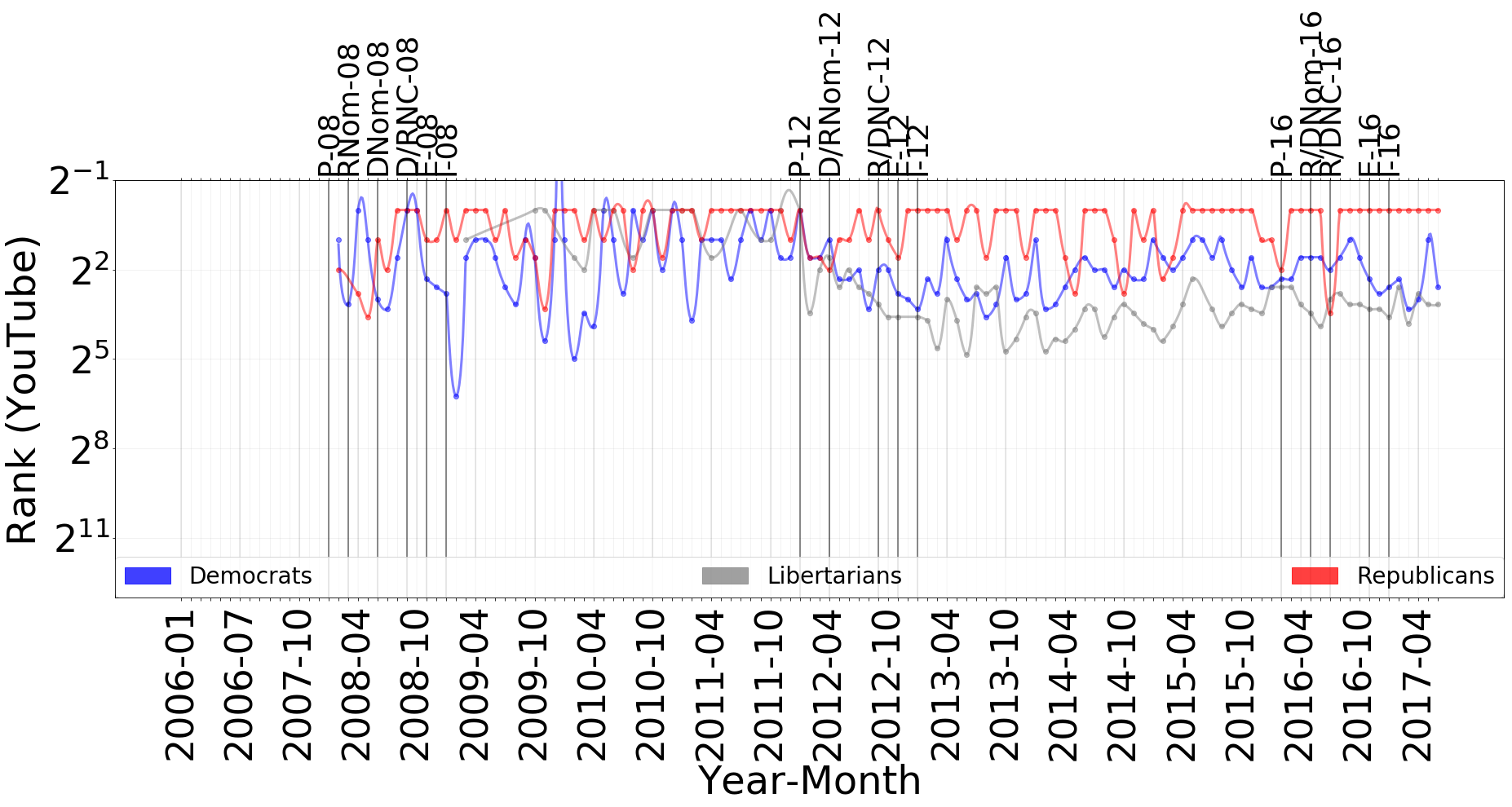}
\caption{ YouTube}
\label{fig:media:yt-rank}
\end{subfigure}

\begin{subfigure}[b]{.49\textwidth}
\includegraphics[trim=0cm 0cm 0cm 0cm, clip=true, width=\textwidth, angle=0]
{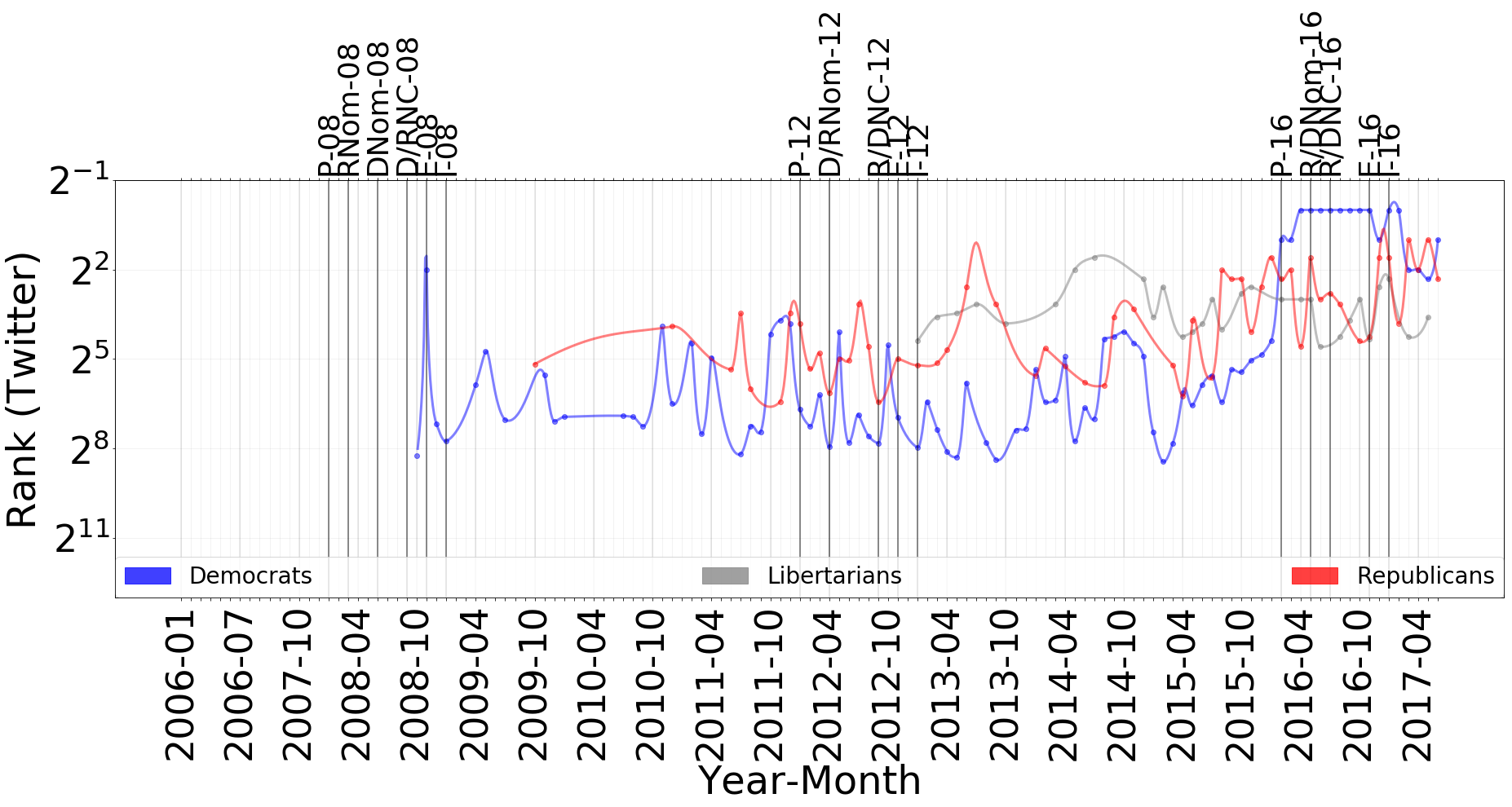}
\caption{Twitter}
\label{fig:media:tw-rank}
\end{subfigure}

\begin{subfigure}[b]{.49\textwidth}
\includegraphics[trim=0cm 0cm 0cm 0cm, clip=true, width=\textwidth, angle=0]
{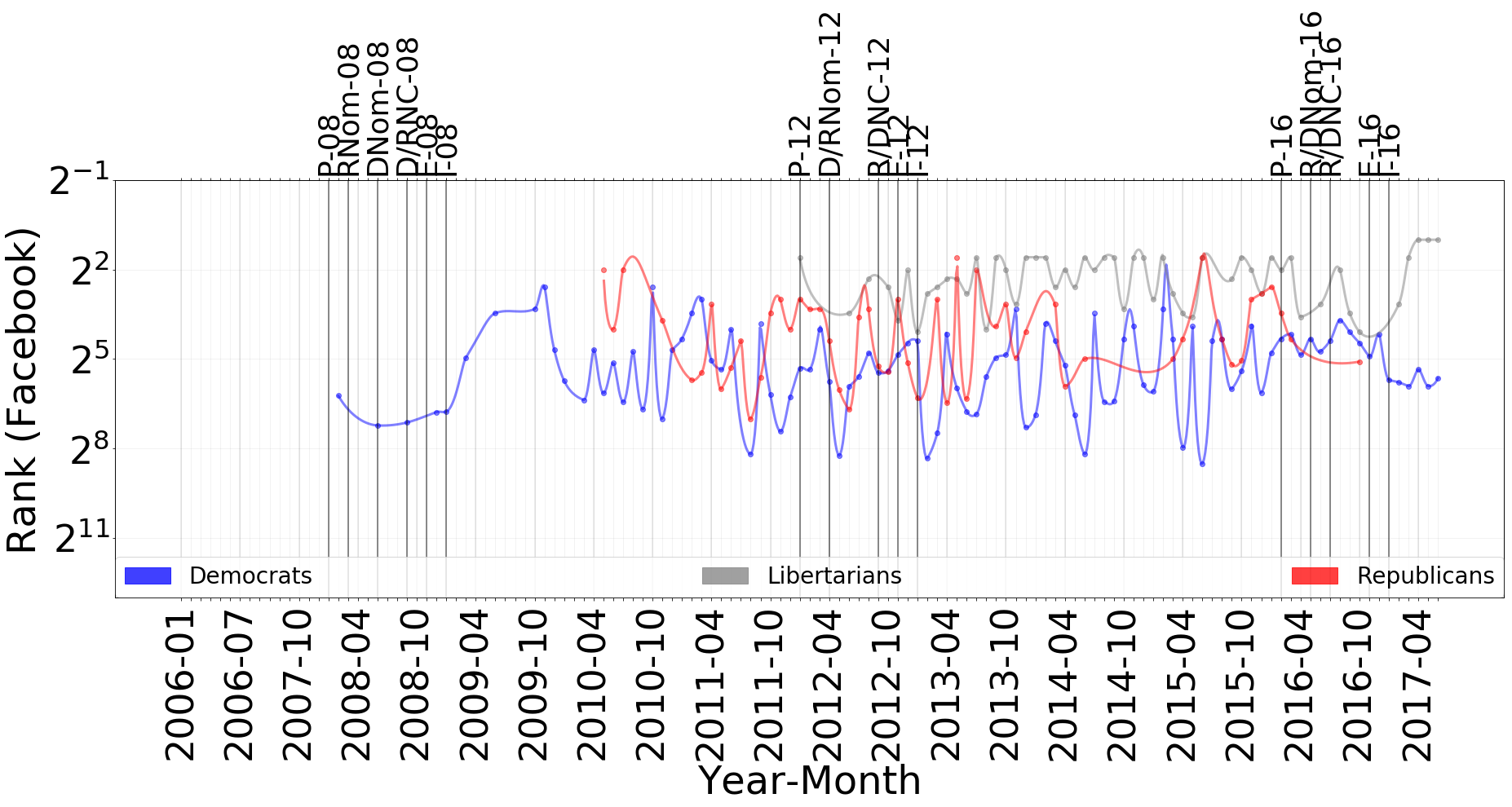}
\caption{Facebook}
\label{fig:media:fb-rank}
\end{subfigure}

\caption{Ranking (by number of comments generated) of YouTube, Twitter, and
Facebook among posts from all media platforms in partisan subreddits.}
\label{fig:media}
\end{figure}

\myparab{Social platforms as news sources.}
Recent polls by Gallup \cite{gallup-media} have shown that trust in traditional
media sources is at an all time low and is continuing to decline.
Simultaneously, the 2016 US presidential election witnessed an explosion of
political discourse on social and democratized media platforms -- particularly
YouTube, Twitter, and Facebook. This is confirmed by  
\Cref{fig:media} which shows how YouTube, Twitter, and Facebook have risen to
prominence as top sources of discussion and information in political subreddits.
The changing landscape of media consumption for politics is apparent. 
When ranked by amount of discussion generated (in terms of comments posted), we
find that each category of subreddits has a preferred social media platform.
Republican subreddits have used YouTube as their top information source since
the 2008 US presidential elections. Since this time, YouTube has been ranked in
the top 10 media outlets for all but five months. In fact, it remained ranked
number one all through the period since the conclusion of the Republican
National Convention until May 2017. On the Democratic and Libertarian
subreddits, we see that YouTube only occasionally appears within the top 10
media outlets. Instead we see that Twitter was the top source of discussion on
the Democratic subreddits for the period since Super Tuesday I until Election
day. Interestingly, unlike the Republican affinity for
YouTube which has been constantly high since 2009, the Democratic affinity for
Twitter increased drastically during the primaries. We find that Republican
subreddits are also increasingly using Twitter as a source of information with
the site moving into and staying in the top 5 ranks since February 2017.
Facebook does not appear to have a major impact in Democratic and Republican
subreddits -- only occasionally entering the top 20 ranks. However, Libertarian
subreddits have consistently had Facebook amongst their top 15 media outlets
since the start of the 2012 election cycle. Since the conclusion of the 2016
elections, Facebook has become the top information source for Libertarian
subreddits.

In terms of impact on political discourse, we find statistically significant
negative  correlations between the incidence of posts from social platforms and
the complexity of discourse, both in the Democratic (Pearson correlation
co-efficient: -.32, p-value < .0005) and Republican (Pearson correlation
co-efficient: -.64, p-value < .0001) subreddits. When considering all political
subreddits, a similar negative  correlation was found (Pearson correlation
co-efficient: -.31, p-value: < .001). No statistically significant correlations
were found when considering the offensiveness of political discourse.
 
\emph{Takeaway:} Posts linking to social media platforms generated significant
amounts of activity in subreddits associated with all parties during the 2016
elections -- Democratic subreddit engagement with posts from Twitter reached a
historical high, Republican subreddits continued to show strong preference for
posts linking to videos on YouTube, and Libertarian affinity for posts linking
to Facebook pages continued to grow. Social media posts have a moderate negative
correlation on the complexity of discourse.
 
\section{Fringe Groups in the Mainstream}\label{sec:fringe}

Recent events -- \eg Unite the Right and White Nationalist rallies across the
country and the Anti-Fascist rallies in response to them -- have shown that
fringe groups and extremists have now infiltrated mainstream political discourse
in the real world. In this section we investigate their participation in
mainstream political subreddits. To measure of influence of an extremist group
we identify redditors that are simultaneously active in at least one hate
subreddit and one political subreddit. We say that a redditor is \emph{active}
in a subreddit for a given month if they have at least (1) 10\% of their monthly
total of comments or posts or (2) at least 10 posts or comments in a subreddit
for a given month. Our list of hate subreddits include 274 (banned, quarantined,
and still open) subreddits associated with racism -- \eg \emph{r/coontown} and
\emph{r/nazi}, sexism -- \eg \emph{r/TheRedPill} and
\emph{r/mensrights}, violence -- \eg \emph{r/killingwomen} and
\emph{r/beatingtrannies}, and peddling conspiracy theories and fake news -- \eg
\emph{r/conspiracy} and \emph{r/blackcrime}. The list of subreddits was gathered
through mining comments from \emph{r/againsthatesubreddits} and announcements of
subreddit bans and quarantines. We note that 87 of our 274 hate subreddits
have been active for over 5 years and that 218 were active even prior to the
start of the 2016 US presidential election season (May 2015).

\begin{figure}[htb]
\centering
\begin{subfigure}[b]{.485\textwidth}
\includegraphics[trim=0cm 0cm 0cm 0cm, clip=true, width=\textwidth, angle=0]
{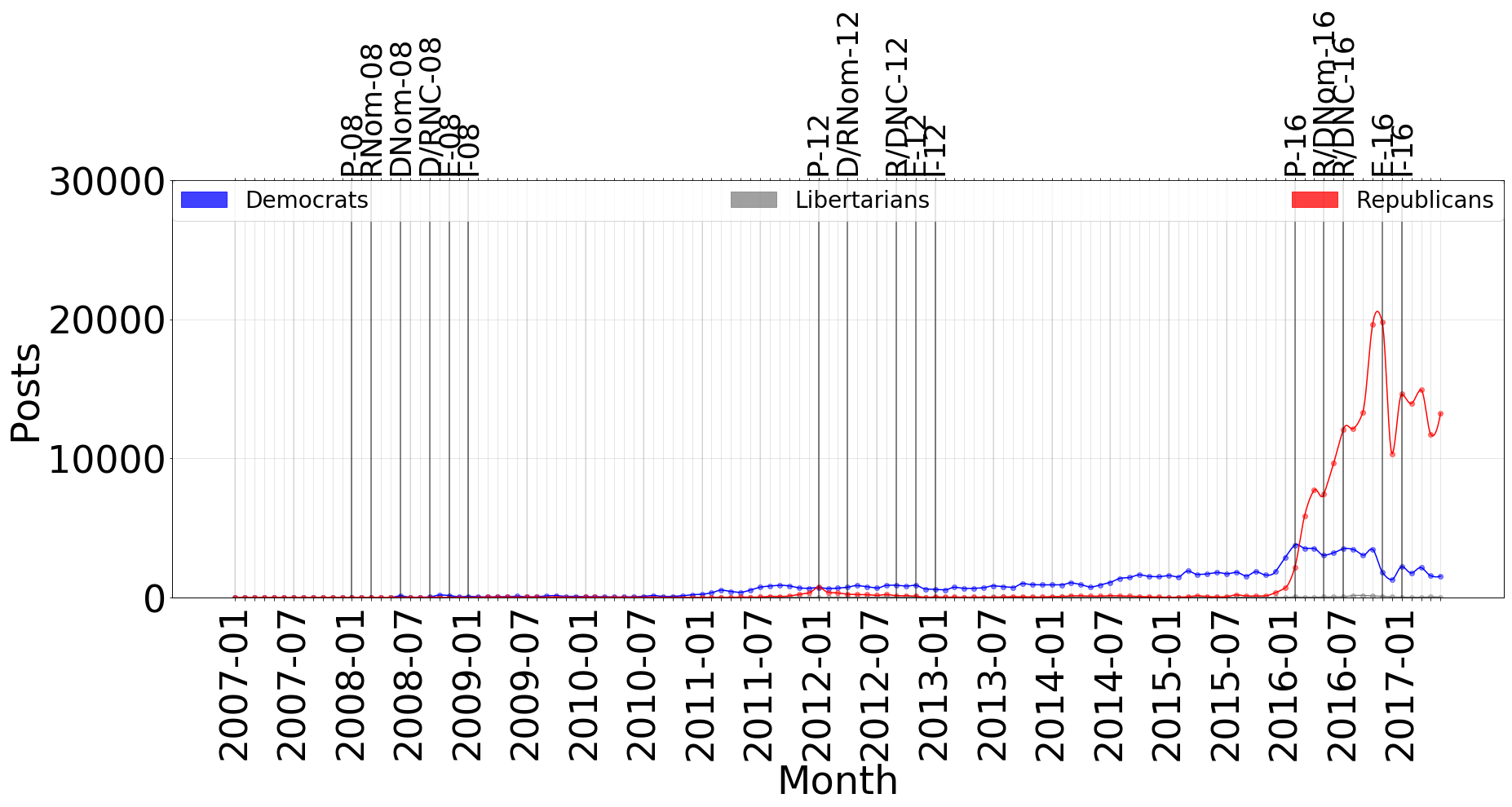}
\caption{Number of posts in partisan subreddits by redditors who were also
active in a hateful subreddit.}
\label{fig:fringe:posts}
\end{subfigure}

\begin{subfigure}[b]{.485\textwidth}
\includegraphics[trim=0cm 0cm 0cm 0cm, clip=true, width=\textwidth, angle=0]
{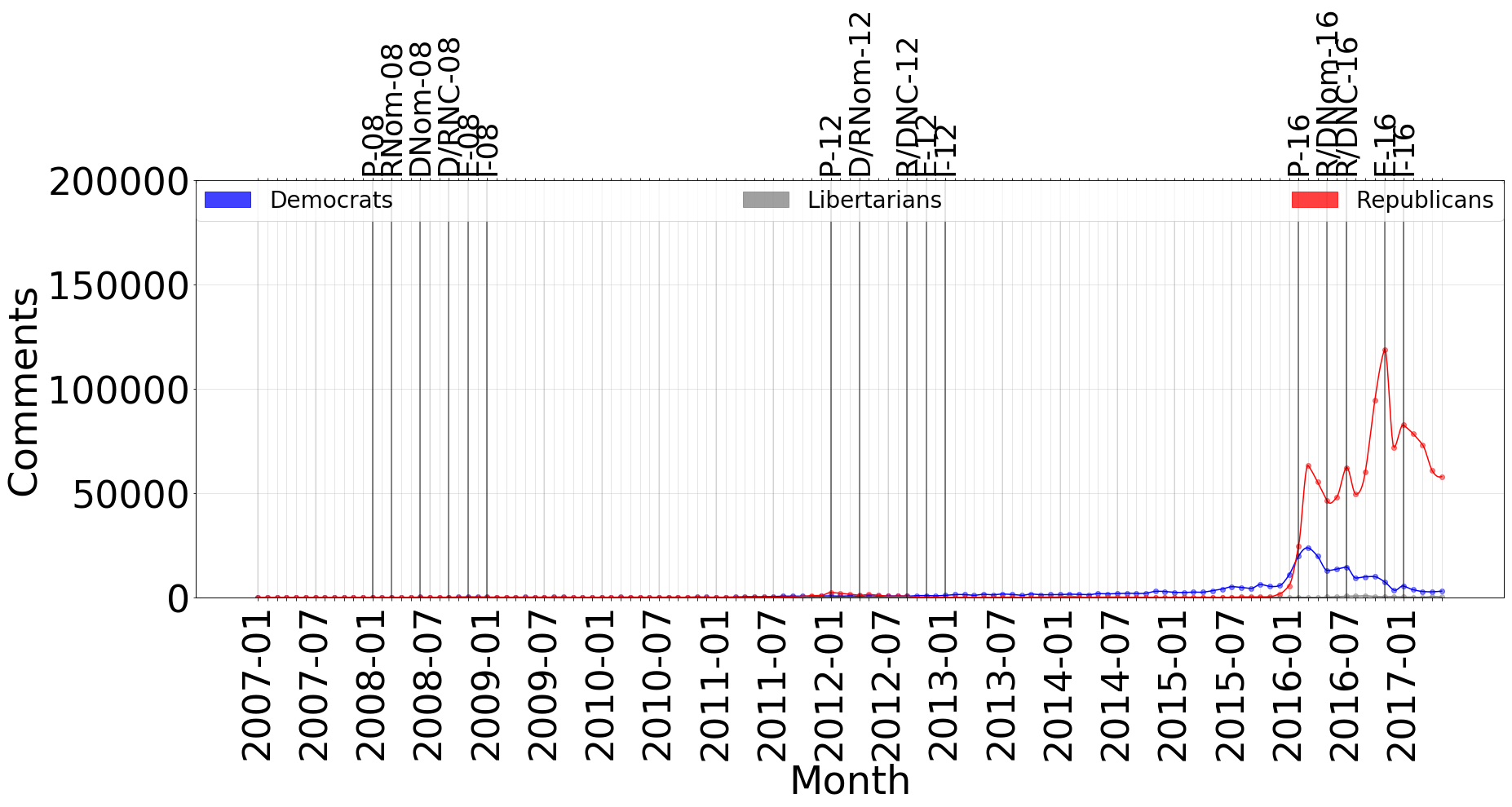}
\caption{Number of comments in partisan subreddits by redditors who were also
active in a hateful subreddit.}
\label{fig:fringe:comments}
\end{subfigure}

\begin{subfigure}[b]{.485\textwidth}
\includegraphics[trim=0cm 0cm 0cm 0cm, clip=true, width=\textwidth, angle=0]
{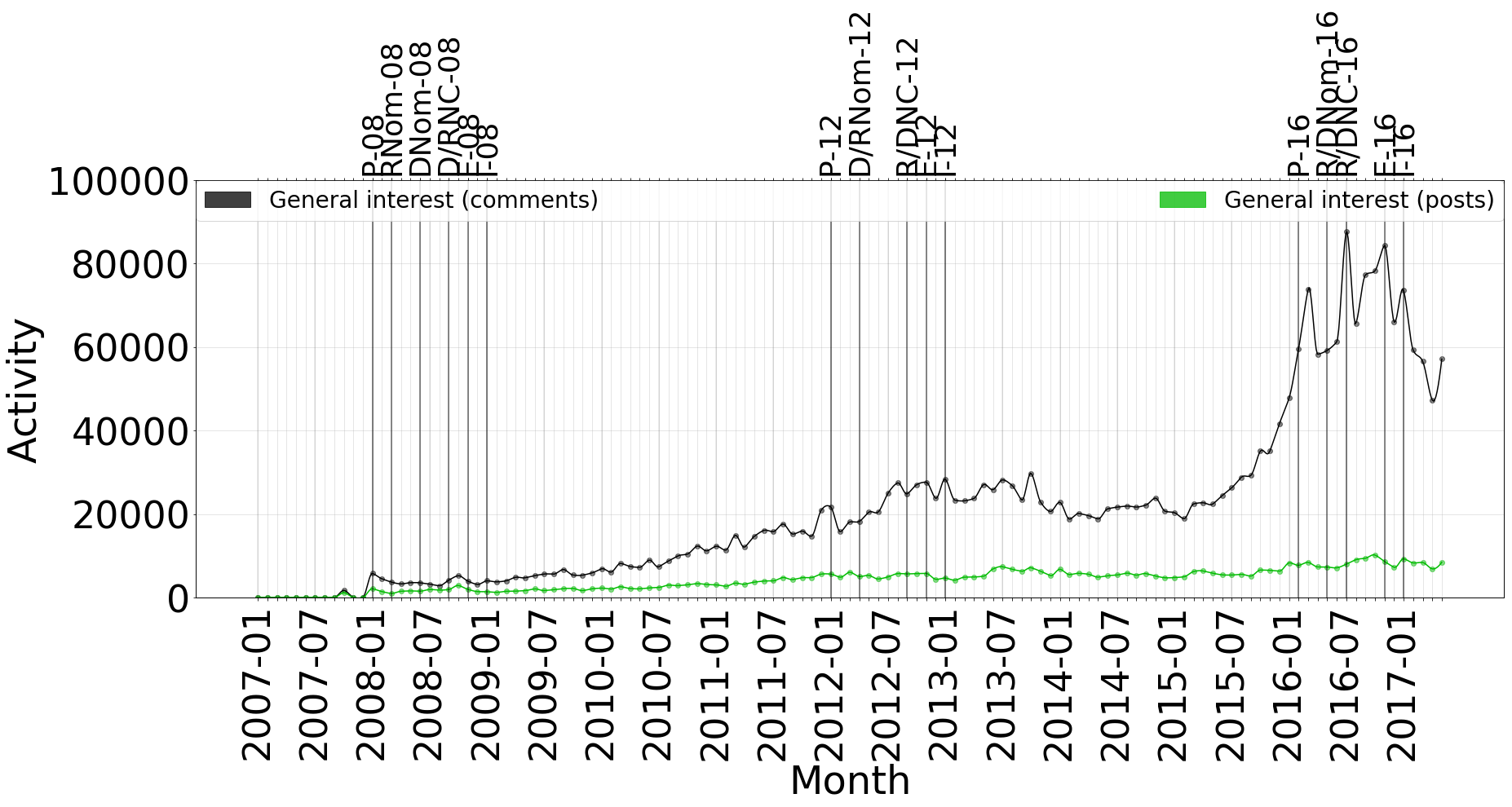}
\caption{Number of comments and posts in non-partisan political subreddits by
redditors who were also active in a hateful subreddit.}
\label{fig:fringe:general}
\end{subfigure}

\caption{Infiltration of fringe (hateful) groups into mainstream political
subreddits.}
\label{fig:fringe}
\end{figure}

\Cref{fig:fringe} shows the number of posts and comments made on mainstream
(partisan and non-partisan) political subreddits by redditors who were, by our
definition, active in an extremist subreddit. We see a startling rise in the
number of posts made by members of extremist subreddits in the partisan and
non-partisan political subreddits. From the period starting in December 2015 and
continuing to Election day, there was a 200\% increase in the number of posts
made by fringe authors in Democratic subreddits and a 6600\% increase in the
number of posts made by these authors in Republican subreddits! At their peak in
November 2016, these authors accounted for over 9\% and 14\% of all posts in
Democratic and Republican subreddits, respectively. Since November, however, we
see that both have declined. As of May 2017, activity of fringe authors on
Democratic subreddits has returned to the pre-election levels, while Republican
subreddits continue to experience 3900\% more posting activity from fringe
redditors (compared to December 2015). Further analysis reveals that 
the fringe subreddits contributing the most to Republican subreddits, in terms
of posting and comment activity, are \emph{r/conspiracy}, \emph{r/TheRedPill},
\emph{r/KotakuInAction},and \emph{r/mensrights}. During November 2016, we found
that over 40\% of all active posters in the following subreddits were
simultaneously active on \emph{r/The\_Donald} -- \emph{r/metacanada} (a
right-wing extremist Canadian subreddit), \emph{r/whiterights},
\emph{r/physical\_removal} (a recently banned subreddit promoting violence
against ``liberals'') and \emph{r/new\_right}. We observe similar overlaps even
in non-partical \emph{general-interest} political subreddits.

We find strong statistically significant correlations between the number of
comments and posts by fringe authors and the levels of offensiveness in
political discourse for partisan and non-partisan subreddits. On the Democratic
subreddits there was a very strong positive correlation (Pearson correlation
co-efficient: .81, p-value < .0001), while there correlation was slightly weaker
on the Republican (Pearson correlation co-efficient: .58, p-value < .0001) and
non-partisan (Pearson correlation co-efficient: .73, p-value < .0001)
subreddits. We found that on Republican subreddits experienced a reduction in
complexity of discourse that was moderately correlated with the increasing
participation from fringe authors (Pearson correlation co-efficient: -.56,
p-value < .0001).

\emph{Takeaway:} At the height of the 2016 presidential elections, Republican
subreddits saw an order of magnitude more activity from active members of
extremist subreddits, while Democratic subreddits saw activity from these
authors double. Since the election, these authors have continued to participate
heavily in Republican subreddits. This infiltration is positively correlated
with the rise of offensiveness in all political discourse.

\section{Discussion}\label{sec:conclusions}

Our investigation of the nature of discourse on Reddit over the past decade has
yielded important insights about how increasing affective partisanship has
influenced the civility of online political discussions. 

First, political
discussions have become substantially more offensive in nature since the launch
of the general election campaign for president in July 2016. Notably, this rise
in incivility is overwhelmingly located on Republican (rather than Democratic)
subreddits. This pattern is consistent with other research that suggests that
polarization has largely been asymmetric, with Republicans exhibiting much more
extremity than Democrats \cite{grossmann2016asymmetric}. 
Second, our analysis suggests that the substantial increase in incivility on
reddit was strongly correlated to the rise of Donald Trump, negative
partisanship, and the mainstreaming offringe groups. When Trump was 
performing well in the polls, incivility also increased, suggesting that his
ascendancy either (1) elicited strong negative reactions from his opponents or
(2) emboldened his supporters, even emboldening holders of extremist ideologies.
Negative partisanship was especially evident during the general election
campaign, as Trump's increasing success elicited more offensive rhetoric in
Democratic subreddits while increasing poll results for Clinton were associated
with more offensive remarks on the Republican side. Research on negative
partisanship predicts that anger will increase when the opposing party is doing
well \cite{mason2016cross, huddy2015expressive}, something we see play out
clearly on reddit during the general election campaign. 
Third, to further analyze the role of negative partisanship, we examined the
sentiments of comments that targeted either party. We find that negative
partisanship continues to grow on Republican subreddits but that it has ebbed a
bit on Democratic subreddits since the 2016 election. On one hand, this runs
counter to what we might expect, as it is usually partisans from the losing
party who react to an election outcome with anger. On the other hand, this fits
with the research suggesting that Republicans generally express higher levels of
negative partisanship than Democrats \cite{iyengar2012affect}. Furthermore, it
may signal the unique nature of Trump's presidency. Specifically, as Trump took
office without winning the popular vote and has constantly been under criticism
since his inauguration, it may not be particularly surprising that the
Republican base feels that their party's status (and the legitimacy of Trump's
presidency) is under threat. This would explain why negative partisanship has
remained high, even as Republicans control both branches of the federal
government. 

Ultimately, 
we are able to demonstrate another unfortunate consequence of America's
political polarization -- namely, the fact that online political discussions have
become remarkably less civil and complex. While these trends are disturbing, we do provide
some reason for hope that the situation can improve. After all, much of our
evidence suggests that the degradation in discourse is tied to the rise of
Trump. Thus, it is possible that our political discussions may become less
offensive when his presence in the limelight fades. 

\bibliographystyle{ACM-Reference-Format}
\bibliography{bibliography} 

\end{document}